\documentclass[prd,aps,10pt,twocolumn,floatfix,nofootinbib,superscriptaddress]{revtex4-1}
\usepackage{lipsum}
\usepackage{textcomp}
\usepackage{amsmath}
\usepackage{amsfonts}
\usepackage{amssymb}
\usepackage{graphicx}
\usepackage{hyperref}
\usepackage{subfigure}
\usepackage[usenames,dvipsnames,svgnames,table]{xcolor}
\usepackage{color}
\usepackage{xcolor}
\usepackage{comment}
\usepackage{mathrsfs}
\usepackage{gensymb}

\begin{document}

\newcommand{\beq}{\begin{equation}}
\newcommand{\eeq}{\end{equation}}
\newcommand{\beqn}{\begin{eqnarray}}
\newcommand{\eeqn}{\end{eqnarray}}
\newcommand{\pa}{\partial}
\newcommand{\vp}{\varphi}
\newcommand{\varep}{\varepsilon}
\newcommand{\ep}{\epsilon}
\newcommand{\br}{{\mbox{\boldmath$r$}}}
\newcommand{\rmH}{{\rm H}}
\newcommand{\rmHe}{{\rm He}}
\newcommand{\rmC}{{\rm C}}
\newcommand{\rmBe}{{\rm Be}}
\newcommand{\rmB}{{\rm B}}
\newcommand{\rmO}{{\rm O}}
\newcommand{\rmLi}{{\rm Li}}
\newcommand{\rmFe}{{\rm Fe}}

\title{Magnetospheres of black hole-neutron star binaries} 

\author{Federico Carrasco}
\email{federico.carrasco@aei.mpg.de}
\affiliation{Max Planck Institute for Gravitational Physics (Albert Einstein Institute), 14476 Potsdam, Germany}
\affiliation{ Instituto de F\'i{}sica Enrique Gaviola, CONICET, Ciudad Universitaria, 5000 C\'o{}rdoba, Argentina}
\author{Masaru Shibata}
\email{masaru.shibata@aei.mpg.de }
\affiliation{Max Planck Institute for Gravitational Physics (Albert Einstein Institute), 14476 Potsdam, Germany}
\affiliation{Center for Gravitational Physics, Yukawa Institute for Theoretical Physics, Kyoto University, 606-8502 Kyoto, Japan}
\author{Oscar Reula}
\email{reula@famaf.unc.edu.ar}
\affiliation{ Instituto de F\'i{}sica Enrique Gaviola, CONICET, Ciudad Universitaria, 5000 C\'o{}rdoba, Argentina}

\date{\today}

\begin{abstract}
We perform force-free simulations for a neutron star orbiting a black hole, aiming at clarifying the main magnetosphere properties of such binaries towards their innermost stable circular orbits. 
Several configurations are explored, varying the orbital separation, the individual spins and misalignment angle among the magnetic and orbital axes.
We find significant electromagnetic luminosities, $L\sim 10^{42-46} \, [B_{\rm pole}/ 10^{12}{\rm G}]^2  \, {\rm erg/s}$ (depending on the specific setting), 
primarily powered by the orbital kinetic energy, being about 
one order of magnitude higher than those expected from unipolar induction. 
The systems typically develop current sheets that extend to long distances following a spiral arm structure. The intense curvature of the black hole produces extreme bending on a particular set of magnetic field lines as it moves along the orbit, leading to magnetic reconnections in the vicinity of the horizon. For the most symmetric scenario (aligned cases), these reconnection events can release large-scale plasmoids that carry the majority of the Poynting fluxes. On the other hand, for misaligned cases, a larger fraction of the luminosity is instead carried outwards by large-amplitude Alfv{\'e}n waves disturbances. We estimate possible precursor electromagnetic emissions based on our numerical solutions, finding radio signals as the most promising candidates to be detectable within distances of $\lesssim 200$\,Mpc by forthcoming facilities like the Square Kilometer Array.

\end{abstract}

%\keywords{}

\maketitle

%%%%%%%%%%%%%%%%%%%%%%%%%%%%%%%%%%%%%%%%%%%%%%%%%%%%%%%%%%%%%%%%%%%%%%%%%%%%%%%%%%%%%%%%%%%%%%%%%%%%%%%%%%%%%%%%%%%%%%%%%%%%%%%%%%%
\section{Introduction}
%%%%%%%%%%%%%%%%%%%%%%%%%%%%%%%%%%%%%%%%%%%%%%%%%%%%%%%%%%%%%%%%%%%%%%%%%%%%%%%%%%%%%%%%%%%%%%%%%%%%%%%%%%%%%%%%%%%%%%%%%%%%%%%%%%%

The spectacular combined detection of gravitational waves (GW) and electromagnetic (EM) radiation from the binary neutron star (BNS) merger GW170817 \cite{abbott17a,abbott17b,abbott17c} has initiated the new era of multi-messenger astronomy. This single multi-messenger observation led to important breakthroughs in our understanding of astrophysics and even fundamental physics.
It confirmed that BNS mergers give rise to short gamma-ray bursts and the production of heavy elements via r-process; as well as it has provided constraints on proprieties of matter at nuclear densities, an estimation of the Hubble constant, and stringent tests of general relativity, among others (e.g., \cite{burns2019}).  

The presence of matter and its support for strong magnetic fields make binary systems involving at least one neutron star (NS) the most likely sources for such simultaneous detection of GW and EM signals. 
The interaction of the NS magnetosphere with its companion during the inspiral phase is expected to generate EM emissions. As they would be produced within the relatively cleaner environment preceding the merger, these precursor EM counterparts --not yet detected-- %or at least not linked to GWs events--, 
could provide crucial information about the sky localization of the source and other physical parameters of the system which cannot be accurately obtained from the GW observation alone.
For black hole-neutron star (BHNS) binaries these precursor EM signals could be particularly relevant, since in most cases these systems are not expected to produce strong post-merger EM emissions: it is known for high enough binary mass-ratios ($\gtrsim 5$) and for a slowly spinning black hole (BH), the NS is likely to be swallowed by the BH without tidal  disruption (e.g., \cite{shibata2015numerical}), and thus, not favoring the ejection of material from the system nor the formation of an accretion disk to support further EM counterparts. In such cases, the existence of precursor EM signals could be the only way to distinguish a BHNS from a binary BH system of the same masses. 

A NS in a compact binary system at late stages prior to merger is likely to be surrounded by a tenuous plasma, well described by the force-free (FF) approximation \cite{gruzinov99}. 
The arguments around this assumption date back to the pioneering work of Goldreich \& Julian in the context of pulsars \cite{goldreich1969}, recently adapted to binaries scenarios in \cite{wada2020}.
Such plasma environment enclosing the binary would allow to tap kinetic energy from the orbital motion (or from individual spins) and, eventually, a part of this EM energy will be re-processed within the magnetosphere leading to emissions on different bands of the EM spectrum.
These central ideas have been well established for isolated pulsars (e.g., \cite{goldreich1969, contopoulos1999, mckinney2006relativistic, timokhin2006force, spitkovsky2006}) and spinning BHs in active galactic nuclei (e.g., \cite{Blandford, Komissarov2004b}).

A few theoretical models have been invoked to describe the energy extraction mechanisms in binary scenarios. One of them focuses on the EM energy-loss produced by the orbital motion of the NS magnetic dipole moment \cite{ioka2000}.
The other focuses on the interaction of the companion moving across the magnetized medium surrounding the NS, typically modeled by the unipolar induction (UI) effect. This idea was introduced long time ago for moving conductors such as satellites around a planet \cite{1965drag, goldreich1969} and later applied to compact binaries \cite{hansen2001,lyutikov2011electro, mcwilliams2011, lai2012dc,piro2012,d2013big}, with the BH acting as a battery in a DC circuit with the NS for the case of a BHNS binary.

Several numerical studies have shown how, indeed, a fraction of the available kinetic energy of compact binary systems can be transferred into the surrounding FF plasma. General relativistic (GR) simulations matching magnetohydrodynamics stellar interiors with an exterior FF magnetosphere were performed for BNS \cite{palenzuela2013electromagnetic, palenzuela2013linking, ponce2014} and, more recently, for BHNS binaries \cite{east2021}. A similar GR FF approach, but assuming a helical Killing vector field, has been carried for the BHNS scenario as well \cite{paschalidis2013}. On the other hand, special relativistic simulations have focused on the energy extraction associated with magnetic dipole radiation by the NS orbital motion \cite{carrasco2020} and the flaring events associated with the twisted magnetic flux-tube arising from the relative motion of two NSs \cite{most2020}.  

This paper aims at further clarifying the main magnetospheric properties of BHNS binaries in their last orbits up-to the innermost stable circular orbit of the system. We assume that the NS is endowed with a dipolar magnetic field and moving  around the BH through an FF plasma environment. 
Instead of performing full GR simulations, we approximate the problem 
by %considering the motion of the star on 
a fixed Kerr spacetime.  Thus, while we are neglecting the curvature of the NS, we are prescribing its trajectory over the BH background. The former assumption might be acceptable acknowledging the dominant dynamical role of the boundary conditions, i.e., anchoring the magnetic field into the NS crust. While the latter is based on the fact that binary trajectories would be essentially determined by the energy-loss due to GWs, which are energetically predominant over any realistic EM interaction \cite{ioka2000, lai2012dc}. 
This simplified setting allows us to conduct rather inexpensive, very accurate, numerical simulations for a detailed description of the magnetospheric response and to explore different relevant configurations and parameters.

In this way, we extend our previous special relativistic study \cite{carrasco2020} to investigate how the intense curvature of the BH affects the resulting BHNS magnetosphere. 
In this paper, we focus only on the binary in circular orbits at several separations close to the innermost stable circular orbit. 
We first consider the case in which the magnetic moment is aligned with respect to the orbital angular momentum and both compact objects are non-spinning. Then, we independently incorporate their spins (aligned with the orbital axis), and later we consider various inclinations of the dipole magnetic moment.   
Besides providing estimations for the amount of kinetic energy that can be extracted from the binary as Poynting luminosity (and/or dissipation) on each setting, we assess how the EM energy is distributed within the magnetosphere and pay particular attention to, e.g., the formation of current sheets (CSs). 
Finally, relying on recent progresses in pulsar theory (e.g., \cite{bai2010, uzdensky2013, kalapotharakos2018, philippov2018, philippov2019pulsar}), we infer possible EM precursor emissions from the attained numerical solutions.

The article is organized as follows. In Sec.~II we setup the problem and describe our numerical implementation. Results are presented in Sec.~III, first focusing on irrotational and aligned BHNS binaries magnetospheres, and then incorporating the effects of spin and misalignment. Possible observational implications of our results are also discussed. We summarize and conclude in Sec.~IV. 
Throughout this paper, unless otherwise stated, we use the geometrical units of $G=1=c$, where $G$ is the gravitational constant and $c$ is the speed of light. 
The Latin indexes $i,j$, and $k$ denote spatial components while other 
Latin indexes ($a,b,c,..$) the spacetime ones. 

%%%%%%%%%%%%%%%%%%%%%%%%%%%%%%%%%%%%%%%%%%%%%%%%%%%%%%%%%%%%%%%%%%%%%%%%%%%%%%%%%%%%%%%%%%%%%%%%%%%%%%%%%%%%%%%%%%%%%%%%%%%%%%%%%%%%% 
\section{Setup}
%%%%%%%%%%%%%%%%%%%%%%%%%%%%%%%%%%%%%%%%%%%%%%%%%%%%%%%%%%%%%%%%%%%%%%%%%%%%%%%%%%%%%%%%%%%%%%%%%%%%%%%%%%%%%%%%%%%%%%%%%%%%%%%%%%%%%

\subsection{General Setting}

As discussed above, we want to consider a star  that is following certain trajectory over a Kerr background. 
The Kerr metric is parametrized by the mass $M$ and spin $a$, and  can be written in the Kerr-Schild form as $g_{ab} = \eta_{a b} + H \, \ell_{a } \ell_{b } $, where $\eta_{ab}$ is the flat metric and $\ell_{a}$ is a null co-vector with respect to both $\eta_{ab}$ and $g_{ab}$. We shall describe the trajectory of the orbiting NS in the Cartesian coordinates $\{x^a \} = \{t,x,y,z\}$ 
associated with the flat part of the metric\footnote{It is sometimes referred as the Kerr-Schild Cartesian coordinates, or the Kerr-Schild frame (see e.g.~\cite{Visser2007}).}. 
In these coordinates, the metric function $H$ takes the form 
\beqn
&& H = \frac{2 M r^3}{r^4 + a^2 z^2} \\
&& r^2 = \frac{1}{2}(\rho^2 - a^2) + \sqrt{\frac{1}{4}(\rho^2 - a^2)^2 + a^2 z^2}\\
&& \rho^2 = x^2 + y^2 + z^2 
\eeqn
and the co-vector $\ell_{a}$ reads
\beq
 \ell_{a} = \left\lbrace 1, \frac{rx + ay}{r^2 + a^2}, \frac{ry-ax}{r^2 + a^2}, \frac{z}{r} \right\rbrace .
\eeq
$\ell^a$ is often defined by $\eta^{ab}\ell_b$, and with this definition, 
we have $\ell^a \ell_a=0$ and $\ell^k \ell_k=1$. Note that $g^{ab}$ is written as
$\eta^{ab}-H\ell^a \ell^b$. 

In the $3+1$ formulation, the line element is written as
\begin{equation}\label{eq:KS}
 ds^2 = \left( \beta_i \beta^i - \alpha^2 \right) dt^2 + 2 \beta_i \, dx^i \, dt + \gamma_{ij} \, dx^i \, x^j,
\end{equation}
with
\beqn
\alpha&=&{1 \over \sqrt{1+H}},\\
\beta_i&=&H \ell_i, ~~~\beta^i={H \over 1+H }\ell^i,\\
\gamma_{ij}&=&\delta_{ij}+H\ell_i \ell_j, ~~~
\gamma^{ij}=\delta^{ij}-{H \over 1+H} \ell^i \ell^j. 
\eeqn
representing the lapse function, shift vector, and spatial metric, respectively. 

The trajectory is defined by the orbital separation $r_o (t)$ and the phase $\varphi_o (t)$ (being $\Omega_o (t) \equiv \dot{\varphi_o} (t) $ the associated angular velocity).
However, our numerical domain will be centered on the NS instead (as in \cite{carrasco2020}), thus describing the dynamics for an adapted foliation with coordinates $\{ \hat{x}^a \} = \{\hat{t},\hat{x},\hat{y},\hat{z}\}$. 
The coordinates transformation into this ``co-moving" foliation is defined by:
\begin{eqnarray}
 && t = \hat{t}, \quad z = \hat{z}, \nonumber \\
 && x = \hat{x} + r_{o}(\hat{t}) \cos (\varphi_o (\hat{t})), \label{coord-transf} \\
 && y = \hat{y} + r_{o}(\hat{t}) \sin (\varphi_o (\hat{t})),  \nonumber
\end{eqnarray}
and thus,
\begin{eqnarray*}
 && dt = d\hat{t}, \quad dz = d\hat{z},  \\
 && dx = d\hat{x} + \left[ \dot{r}_{o}(\hat{t}) \cos (\varphi_o (\hat{t})) - r_{o}(\hat{t}) \Omega_{o}(\hat{t}) \sin (\varphi_o (\hat{t}) ) \right]  d\hat{t},  \\
 && dy = d\hat{y} + \left[ \dot{r}_{o}(\hat{t}) \sin (\varphi_o (\hat{t})) + r_{o}(\hat{t}) \Omega_{o}(\hat{t}) \cos (\varphi_o (\hat{t}) ) \right]  d\hat{t}.  
\end{eqnarray*}
Therefore, the line element \eqref{eq:KS} in the new coordinates is written simply as
\begin{equation}\label{eq:metric}
 d\hat{s}^2 = \left( \hat{\beta}_i \hat{\beta}^i - \hat{\alpha}^2 \right) d\hat{t}^2 + 2 \hat{\beta}_i \, d\hat{x}^i \, d\hat{t} + \hat{\gamma}_{ij} \, d\hat{x}^i \, d\hat{x}^j,
\end{equation}
with
\begin{eqnarray}
 \hat{\alpha} &=& \alpha(x(\hat{x})), \\
 \hat{\beta}^i &=& \beta^{i}(x(\hat{x})) + \beta_{o}^i, \\
 \hat{\gamma}_{ij} &=& \gamma_{ij}(x(\hat{x})),
\end{eqnarray}
where a new piece of the shift vector due to the orbital motion is
\begin{equation}
 \beta_{o}^i =  \left\lbrace \dot{r}_o  \cos \varphi_o  - v_o \sin \varphi_o , \dot{r}_o  \sin \varphi_o + v_o \cos \varphi_o, 0 \right\rbrace. 
\end{equation}
In the above, to simplify the notation, we have dropped all time dependencies and defined $v_o := r_o \Omega_o$. ``$(x(\hat{x}))$" on the metric components above implies the same functional form of the previous ones, but evaluated at the ``displaced" spatial points as seen in the new coordinate system. 

\subsection{Evolution Equations}

Force-free electrodynamics (FFE) is governed by Maxwell's equations,
\begin{eqnarray}
  \nabla_b F^{ab} &=& j^{a},  \label{Max} \\
  \nabla_b F^{*ab} &=& 0,
\end{eqnarray}
together with the FF condition,
\begin{equation} \label{FF-cond}
 F_{ab}j^{b} = 0. 
\end{equation}
Here, $F^{ab}$ is the EM tensor, $F^{*ab}$ its dual, and $\nabla_b$ the 
covariant derivative with respect to $g_{ab}$. 
In this regime, remarkably, the EM field can be evolved autonomously
without keeping track of the plasma degrees of freedom. 
This is achieved by 
%eliminating the plasma current $j^{a}$ from the description, that is: rewriting \eqref{Max} as
%\begin{equation}
%F_{ab} \nabla_{c} F^{bc} = 0
%\end{equation} 
%or equivalently, 
finding an expression for $j^{a}$ in terms of the EM field. 
%In order to write-down evolution equations for the electric and magnetic fields on a given $3+1$ splitting (for either of these approaches), it is necessary to assume at some point an equation like, $\partial_t (E\cdot B) = 0$, to complete the system~\footnote{This equation provides the necessary information for the evolution of the electric field --or analogously, the component of the current-- along the direction of the magnetic field.
%Essentially, those terms along $B^i$ in the plasma current $j^{a}$ are responsible for the screening of the parallel electric field that guarantees $E_i B^i = 0$. This is the reason why, some authors does not include this terms in the equations and enforce, instead, the force-free condition $E_i B^i = 0$ by other strategies.}.  

Following \cite{shibata2015numerical}, we can write the corresponding evolution equations for this system in a pseudo-conservative form:
\begin{eqnarray}
&& \partial_t \mathcal{E}^i = -\partial_k \mathcal{F}^{ki} - \mathcal{I}^i,  \label{cons-E}\\
&& \partial_t \mathcal{B}^i = -\partial_k \mathcal{F^*}^{ki},    \label{cons-B}
\end{eqnarray}
where $\mathcal{E}^i := \sqrt{\gamma} E^i$ and $ \mathcal{B}^i := \sqrt{\gamma} B^i$ are the densitized variables with $E_a := F_{ab} n^b$ and $B_a := -F^{*}_{ab} n^b$ 
($n^a$ is the unit time-like vector normal to the spatial hypersurfaces), and
their associated fluxes are given by:
\begin{eqnarray}
 \mathcal{F}^{ki} &:=& \alpha \sqrt{\gamma} F^{ki} \text{	} = \beta^i \mathcal{E}^k - \beta^k \mathcal{E}^i + \alpha \epsilon^{kij} \mathcal{B}_j, \\
\mathcal{F^{*}}^{ki} &:=& \alpha \sqrt{\gamma} F^{*ki} = \beta^i \mathcal{B}^k - \beta^k \mathcal{B}^i - \alpha \epsilon^{kij} \mathcal{E}_j,
\end{eqnarray}
with $\epsilon_{abc} = n^{d}\epsilon_{dabc}$ being the induced volume element on the spatial hypersurfaces. The FF current reads,
\begin{eqnarray}\label{cons-I}
 \mathcal{I}^i &=& \mathcal{Q} \left( \frac{ \alpha\mathcal{S}^i}{\mathcal{B}^2} - \beta^i \right) \nonumber \\
               &+& \frac{\mathcal{B}^i}{\mathcal{B}^2} \left[ \mathcal{E}_j \partial_k \mathcal{F^{*}}^{kj} - \mathcal{B}_j \partial_k \mathcal{F}^{kj} + \mathcal{E}^{k} \mathcal{B}^{j} \mathcal{L}_{t} \gamma_{kj}   \right], %\nonumber 
\end{eqnarray}
where $\mathcal{Q}:= \sqrt{\gamma} \rho_e = \partial_k \mathcal{E}^k$ is the electric charge density\footnote{
Notice that in an stationary spacetime one has that $\mathcal{L}_{t} \gamma_{kj} = 0$, although this will not be the case if we are to evolve the system with the co-moving foliation $\{ \hat{x}^a \}$ in which the time vector $\hat{t}^a$ is not a Killing field (as it was the case of $t^a$ in the original Kerr-Schild coordinates $\{ x^a \}$).}. 

However, this formulation of the theory is only weakly hyperbolic \cite{Pfeiffer}.
Therefore, one has to use either a symmetrized version of the theory like the one we developed in \cite{FFE} or a simplified version of the equations without the terms parallel to $ \mathcal{B}^i$ in the current \eqref{cons-I}, i.e.,
\begin{eqnarray}
&& \partial_t \mathcal{E}^i =  -\partial_k \mathcal{F}^{ki} - \mathcal{Q} \left(  \frac{\alpha \mathcal{S}^i}{\mathcal{B}^2} - \beta^i \right),  \\
&& \partial_t \mathcal{B}^i =  -\partial_k \mathcal{F}^{*ki}   ,
\end{eqnarray}
while numerically enforcing $\mathcal{E} \cdot \mathcal{B} = 0$ (like in \cite{spitkovsky2006, Palenzuela2010Mag, alic2012}).
We choose this simplified version of FFE since it is much straightforward to implement and has shown to produce very similar results to those with the more elaborated system \cite{FFE} (this was tested on the similar setting of \cite{carrasco2020}).

\subsection{Initial and Boundary Data}\label{sec2-3}

The initial configuration is set by a magnetic dipolar field and vanishing electric field. We approximate this EM field from its flat spacetime vector potential solution, centered in our adapted coordinates $\{ \hat{x}^a \}$,
\begin{equation}
 A_{\hat{\phi}} = \frac{\mu \sin^2 \hat{\theta}}{\hat{r}},
\end{equation}
where $\mu$ is the magnetic dipole-moment. 
We focus primarily on the cases in which the magnetic axis is aligned with the orbital angular momentum, but we also consider other scenarios in which these two axes are not aligned. 
In our setup, the $\hat{z}$  axis is always perpendicular to the orbital plane, and thus, the misalignment is attained by just tilting the initial magnetic moment by an angle $\chi$ along the $\hat{x}$-$\hat{z}$ plane.

We will consider relevant (fixed) values of the orbital separation $r_o$, setting the orbital angular velocity accordingly as (e.g., chapter 12 of \cite{ShapiroTeukolsky})
\begin{equation}\label{eq:frequency}
 \Omega_o = \frac{(M_{\rm BH}+M_{\rm NS})^{1/2}}{r_{o}^{3/2} + a M_{\rm BH}^{1/2}}.
\end{equation}
Here we incorporate the effect of the BH spin on the orbital motion phenomenologically mimicking the angular velocity for a particle orbiting the spinning BH~\cite{ShapiroTeukolsky}.
The NS is gradually set in motion (at this fixed $r_o$) until it reaches the corresponding orbital frequency \eqref{eq:frequency} %after some time $t=t_0$; 
and, from then on, the system follows its prescribed circular orbit trajectory.  
 
The boundary condition at the NS surface is derived as in \cite{carrasco2020} by assuming the perfectly conducting condition,
\begin{equation}
 0 = F_{ab} \hat{t}^a = F_{ab} ( \hat{\alpha} \hat{n}^a + \hat{\beta}^a ),
\end{equation}
which can be easily generalized to incorporate the NS spin at frequency $\Omega_*$ by,
\begin{equation}
 F_{ab} ( \hat{t}^a + \Omega_* \hat{\phi}^a ) = 0,
\end{equation}
where $\hat{\phi}^a \equiv (\partial_{\hat{\phi}})^a $. Thus, the resulting condition on the electric field measured by a fiducial observer in this adapted foliation (i.e. $\hat{E}_a := F_{ab} \hat{n}^a$) can be written as
\begin{equation}\label{eq:BC}
 \hat{E}^i =  -\frac{1}{\hat{\alpha}} \epsilon^{i}_{\phantom{i}jk} (\hat{\beta}^j + \Omega_* \hat{\phi}^j ) \hat{B}^k .
\end{equation}

\subsection{Numerical Implementation}

We evolve the simplified version of FFE described above, with the coordinate system centered on the NS and with the inclusion of a divergence cleaning field $\phi$, i.e.:
\begin{eqnarray}
\partial_{\hat{t}} \mathcal{E}^{\hat{i}} &=& - \partial_{\hat{k}} \mathcal{F}^{\hat{k}\hat{i}} - \partial_{\hat{k}} \mathcal{E}^{\hat{k}} \left(  \frac{\hat{\alpha} \mathcal{S}^{\hat{i}}}{\mathcal{B}^2} - \hat{\beta}^{\hat{i}} \right),  \\
\partial_{\hat{t}} \mathcal{B}^{\hat{i}} &=& - \partial_{\hat{k}} \mathcal{F^*}^{\hat{k}\hat{i}} + \hat{\beta}^{\hat{i}} \partial_{\hat{k}}  \mathcal{B}^{\hat{k}} - \hat{\alpha} \hat{\gamma}^{\hat{i}\hat{k}} \partial_{\hat{k}} \phi, \\
\partial_{\hat{t}} \phi &=& \hat{\beta}^{\hat{k}} \partial_{\hat{k}} \phi - \hat{\alpha} \partial_{\hat{k}} \mathcal{B}^{\hat{k}} - \hat{\alpha} \kappa \phi .
\end{eqnarray}

Our numerical scheme to solve these equations is based on the \textit{multi-block approach} \cite{Leco_1, Carpenter1994, Carpenter1999, Carpenter2001}, %in which the numerical domain is built from several non-overlapping grids where only grid-points at their boundaries are sheared. The equations are discretized at each individual subdomain by using difference operators constructed to satisfy summation by parts. In particular, we employ difference operators which are eighth-order accurate on the interior and fourth-order at the boundaries. 
%Numerical dissipation is incorporated through the use of adapted Kreiss-Oliger operators. These compatible difference and dissipation operators were both taken from \cite{Tiglio2007}. A fourth order Runge-Kutta method is used for time integration. 
relying on higher-order finite difference operators with an adapted Kreiss-Oliger dissipation \cite{Tiglio2007} and Runge-Kutta method for integration in time. The numerical approach, including the treatment given to boundary conditions and CSs, is exactly the same as in \cite{carrasco2020}; hence, we refer the readers to \cite{carrasco2020} (and to the references therein) for further details. 
In the present version, the FF condition $\mathcal{E} \cdot \mathcal{B} = 0$ is enforced through the projection, $\displaystyle \mathcal{E}^{\hat{i}} \rightarrow \mathcal{E}^{\hat{i}} - \frac{\mathcal{E}\cdot\mathcal{B}}{\mathcal{B}^2} \, \mathcal{B}^{\hat{i}} $,
%\begin{equation}
% \mathcal{E}^i \rightarrow \mathcal{E}^i - \frac{\mathcal{E}\cdot\mathcal{B}}{\mathcal{B}^2} \, \mathcal{B}^i
%\end{equation}
at each Runge-Kutta sub step.

Since the BH interior is contained within our numerical domain, we need to regularize the metric functions inside the outer horizon located in the Boyer-Lindquist coordinates at $r_{+} = M_{\rm BH} + \sqrt{M_{\rm BH}^2 - a^2}$. 
We shall do it by first rewriting $H$ and $\ell_i$ as
\begin{eqnarray}
&& H  \equiv \frac{2M/r}{1+(a/r)^2 (z/r)^2}, \\ %= \frac{2 M r^3}{r^4 + a^2 z^2}
&& \ell_{i}  \equiv \frac{1}{1+(a/r)^2} \left[ \frac{r_i}{r} + (a/r) \frac{v_i}{r} \right],  %= \left\lbrace \frac{rx + ay}{r^2 + a^2}, \frac{ry-ax}{r^2 + a^2}, \frac{z}{r} \right\rbrace
\end{eqnarray}
where $r_i := \left\lbrace x,y,z \right\rbrace$ and $v_i := \left\lbrace y, -x, (a/r) z \right\rbrace$.
Then, we replace $\displaystyle \frac{1}{r} \rightarrow \frac{1}{r_c} p(r/r_{c})$ everywhere for $r<r_c= \lambda r_{+}$ with $\lambda$ being less than unity (typically set to $\lambda = 0.6$), and $p(r/r_{c})$ a large polynomial in $r/r_{c}$ built to guarantee smoothness across the matching surface $r=r_c$. %up-to six derivatives
%\begin{equation*}
%p(r/r_{c}) = 7 - 21 r/r_c + 35 (r/r_{c})^2 - 35 (r/r_{c})^3 + 21 (r/r_{c})^4 - 7 (r/r_{c})^5 + (r/r_{c})^6
%\end{equation*}

Of course, such modifications to the metric inside the BH horizon do not affect the physics outside, since these two regions are causally disconnected. However, we further prevent the production and amplification of high-frequency numerical noises inside the BH by damping the electric and magnetic fields through the following replacements:
\begin{eqnarray}
 && \partial_{\hat{t}} \mathcal{E}^{\hat{i}} \rightarrow \partial_{\hat{t}} \mathcal{E}^{\hat{i}} + \kappa \mathcal{E}^{\hat{i}}  \qquad (r< r_{+}),\\
  && \partial_{\hat{t}} \mathcal{B}^{\hat{i}} \rightarrow \partial_{\hat{t}} \mathcal{B}^{\hat{i}} + \kappa \mathcal{B}^{\hat{i}} \qquad (r< r_{+}),
\end{eqnarray}
with damping parameter $\kappa = 60 M_{\rm BH}^{-1}$ in our simulations.

We solve the system in a region between an interior sphere at radius $\hat{r}=R_*$, which represents the NS surface (i.e., $R_*$ denotes the NS radius), and an exterior spherical surface located at $\hat{r}\approx 128 R_*$.
The domain is represented by a total of $6 \times 7$ sub-domains, with $6$ patches to cover the angular directions and $7$  being the number of spherical shells expanding in radius. These spherical shells do not overlap each other, and have more resolution in the inner regions: from layer to layer, the radial resolution is decreased by a factor $2$.
Typically, we adopt a resolution with total grid numbers of $N_{\hat{\theta}} \times N_{\hat{\phi}} \times N_{\hat{r}}$ with $N_{\hat{\phi}} = 2 N_{\hat{\theta}} = 320$, while $N_{\hat{r}} = 560$ to span the whole computational domain. 
To give an idea, for a binary of mass ratio $q(=M_{\rm BH}/M_{\rm NS})=3$ with orbital separation $r_o = 10 M_{\rm BH}$ at this resolution, the BH is covered by $20 \times 20 \times 35$ grid-points.

\subsection{Analysis Quantities}

We monitor the EM energy and its associated fluxes.  
In FFE the four-momentum, $p^a = -T_{EM}^{ab} t_b$, is conserved (i.e. $\nabla_a p^a = 0$)~\footnote{This equation is satisfied except for regions like CS at which dissipation occurs.} in stationary spacetimes where $t^a \equiv \left( \partial_t\right)^a $ is a Killing field.  
%In the co-moving coordinates $\{ \hat{x}^a \}$, it reads:
%\begin{eqnarray}
% p^{a'} &=& - T_{EM}^{a'b'} t_{b'} = - T_{EM}^{a'b'} \hat{n}_{b'} \nonumber\\
%	&=& \frac{1}{2} ( \hat{E}^2 + \hat{B}^2 ) \, \hat{n}^{a'} - \hat{S}^{a'} .
%\end{eqnarray}
Hence, we measure the corresponding energy and fluxes by
\begin{equation*}
 E(t) := \int_{\Sigma_{t}} \varepsilon \sqrt{\gamma}\, d^{3}x \text{,  }\quad L (t, r) := \oint_{r} \Phi_{\varepsilon} \, \sqrt{-g}  \, d^{2}x, 
\end{equation*}
where the Poynting luminosity $L$ is being integrated on spherical surfaces of radius $r$ around the BH and,
\begin{eqnarray}
\varepsilon &:=& -p^{a} n_{a} = \frac{\alpha}{2} ( E^2 + B^2 ) -S^i \beta_i,\\
%- S\cdot \beta,  \\
\Phi_{\varepsilon}  &:=& p^{a} (dr)_{a} \label{eqn:flux}\\
&=& -( E^2 + B^2 - S_0 / \alpha ) \beta^r + E_0 E^r + B_0 B^r  + \alpha S^{r} , \nonumber
\end{eqnarray}
with $S^{i} := \epsilon^{ijk} E_j B_k$ being the spatial Poynting vector and 
%we have denoted some contractions by, 
$X_0 \equiv X^a \beta_a$.

We also monitor the charge distribution and electric currents present during the dynamics. 
Thus, we look at the FF current density along the magnetic field, as seen by an observer moving in the direction normal to the spatial slices $n^{a}$,
\begin{eqnarray}\label{eq:Jparallel}
 J_{\parallel} &:=& \frac{\mathcal{I}^k B_k}{\sqrt{\gamma} |\vec{B}|} \nonumber \\
& = &-\frac{\rho_e}{|\vec{B}|} B_k \beta^k + \frac{E_k}{\sqrt{\gamma} |\vec{B}|} \partial_j \mathcal{F}^{*kj}  -  \frac{B_k}{\sqrt{\gamma} |\vec{B}|} \partial_j \mathcal{F}^{kj}. ~~
\end{eqnarray}

\section{Results}

We aim at understanding in detail the magnetospheric properties of the steady states of BHNS binary systems at given circular orbits near their innermost stable circular orbit.
We thus extend our previous studies~\cite{carrasco2020}, paying a particular attention to the effect that the intense curvature in the vicinity of the BH has on the magnetosphere: e.g., in the magnetic field topology, the induced electric currents and charge distributions, and in the obtained Poynting luminosity. 
Finally, we elaborate on the implications of these results to potential EM observations.

\subsection{Circular Orbits: general features}\label{sec:aligned}

%%%%%%%%%%%%%%%%%%%%%%%%%%%%%%%%%%%%%%%%%%%%%%
\begin{figure*}[!ht]
\centering{
\includegraphics[scale=0.22]{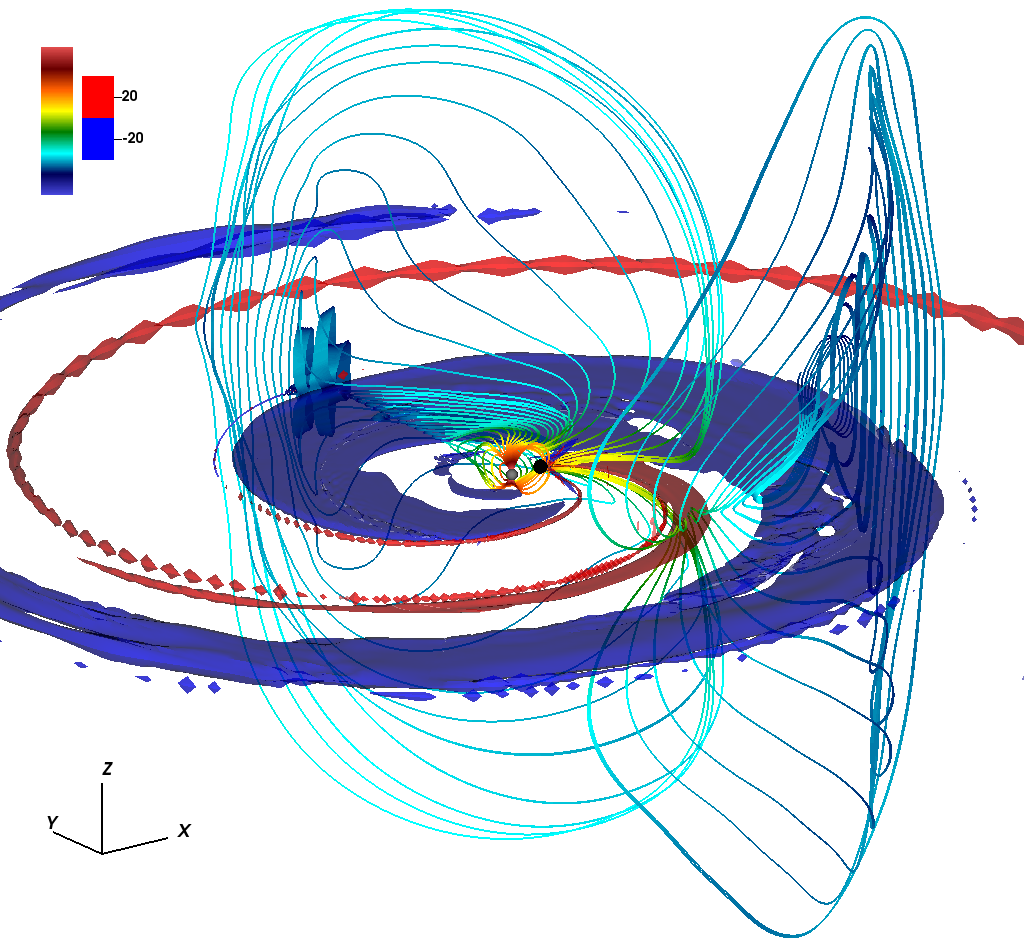}
\includegraphics[scale=0.22]{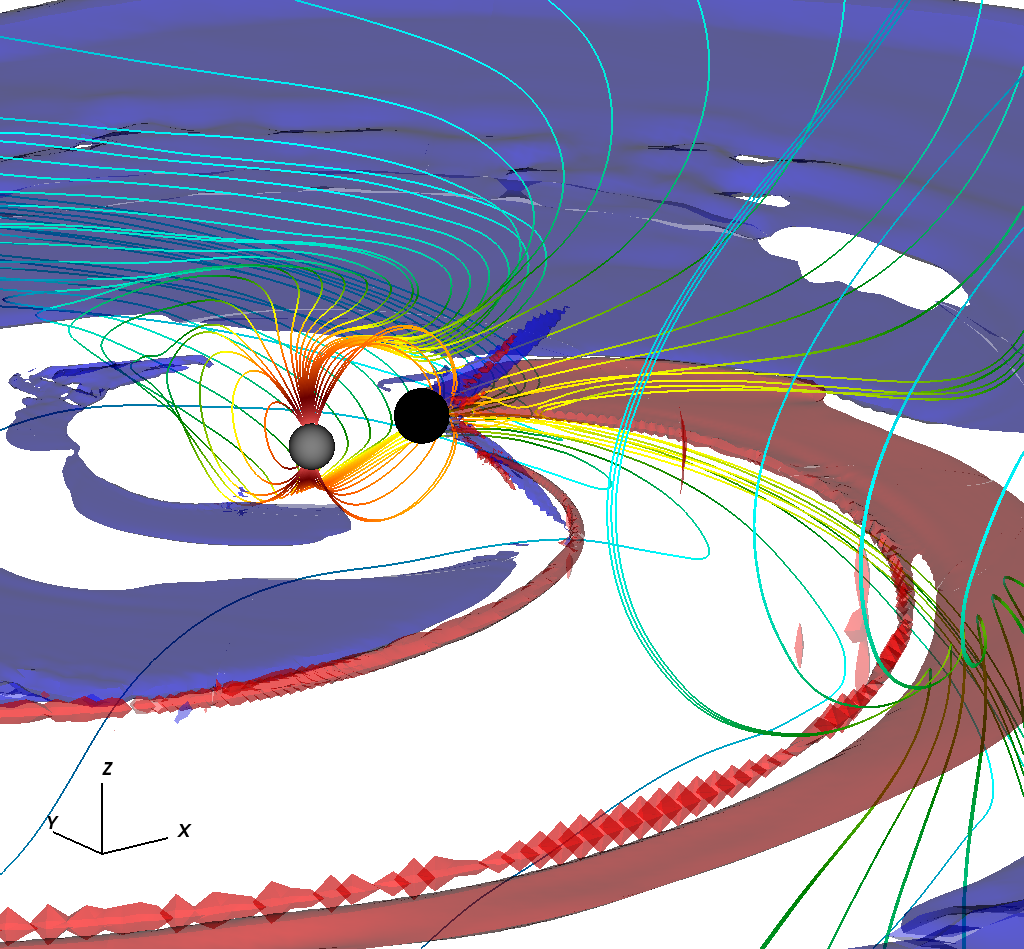}
\caption{
Structure of the magnetosphere, with an emphasis on the bending of magnetic field lines and spiral CS, for a BHNS binary in a circular orbit at an orbital separation $r_o = 10 M_{\rm BH}$ after 3 periods. Left panel: representative magnetic field lines (with color indicating its strength in logarithmic scale);  along with two contours of the electric charge density, normalized by the Goldreich-Julian (GJ) value $\Omega_o B/2\pi c$. Right panel: zoom-in view of the left panel plot, in order to show further details of the BH effect on the magnetic field lines. The black and grey spheres denote the BH horizon and 
NS surface, respectively. 
} 
 \label{fig:B1-CS}}
\end{figure*}
%%%%%%%%%%%%%%%%%%%%%%%%%%%%%%%%%%%%%%%%%%%%%%

We first consider circular orbits at fixed separation $r_o = 10 M_{\rm BH}$, for which both compact objects are non-spinning and the magnetic moment of the NS is aligned to the orbital angular momentum. As mentioned in Sec.~\ref{sec2-3}, 
in the simulations, the NS is gradually set into motion until it attains its final orbital angular velocity given by Eq.~\eqref{eq:frequency}. The system settles into a quasi-stationary regime after about $2$ orbits.
In Fig.~\ref{fig:B1-CS} we illustrate the main features of the solution at $3$ orbital periods, showing both compact objects NS/BH as grey/black surfaces together with carefully chosen magnetic field lines. The contour-plots in the figure represent regions of very large charge density (exceeding the Goldreich-Julian (GJ) values~\cite{goldreich}) signaling the presence of strong CSs. One may distinguish, in principle, two kinds of CSs in the picture. There is a CS that follows a spiral pattern over the orbital plane, associated with  discontinuities of the toroidal magnetic field across the plane, being the sign of the charge density (depicted by the red/blue contours in the plot) indicative of the orientation of this ``jump''. And there is another, out-of-the-plane CS, produced by a discontinuity of the poloidal magnetic field at an outer side of the BH horizon.
It can be seen that the dynamical effect of the curvature is quite strong for a set of magnetic field lines, which bend towards the BH, suffering extreme twisting of about $\sim \pi$ and leading to X-point magnetic reconnections. Such continuous reconnections taking place near the BH produce closed magnetic loops (i.e., plasmoids) that are released at relativistic speeds, transporting electromagnetic energy away from the system\footnote{An animation showing the dynamics can be found at this \href{https://drive.google.com/drive/folders/1CDfiMedswq6ArcUC-NMt4EB0W6DGsmbq?usp=sharing}{link}.}. 
The large-scale plasmoids can be clearly appreciated at the right side of the plots. For the larger loops, produced closer to the BH, the field strengths are higher (as also noticed in \cite{east2021}). On the other hand, the smaller ones, involving weaker magnetic fields, enclose the blue contour component of the equatorial CS (see Fig.~\ref{fig:B1-CS}). 

\begin{figure}%[!ht]
\centering{
\includegraphics[scale=0.18]{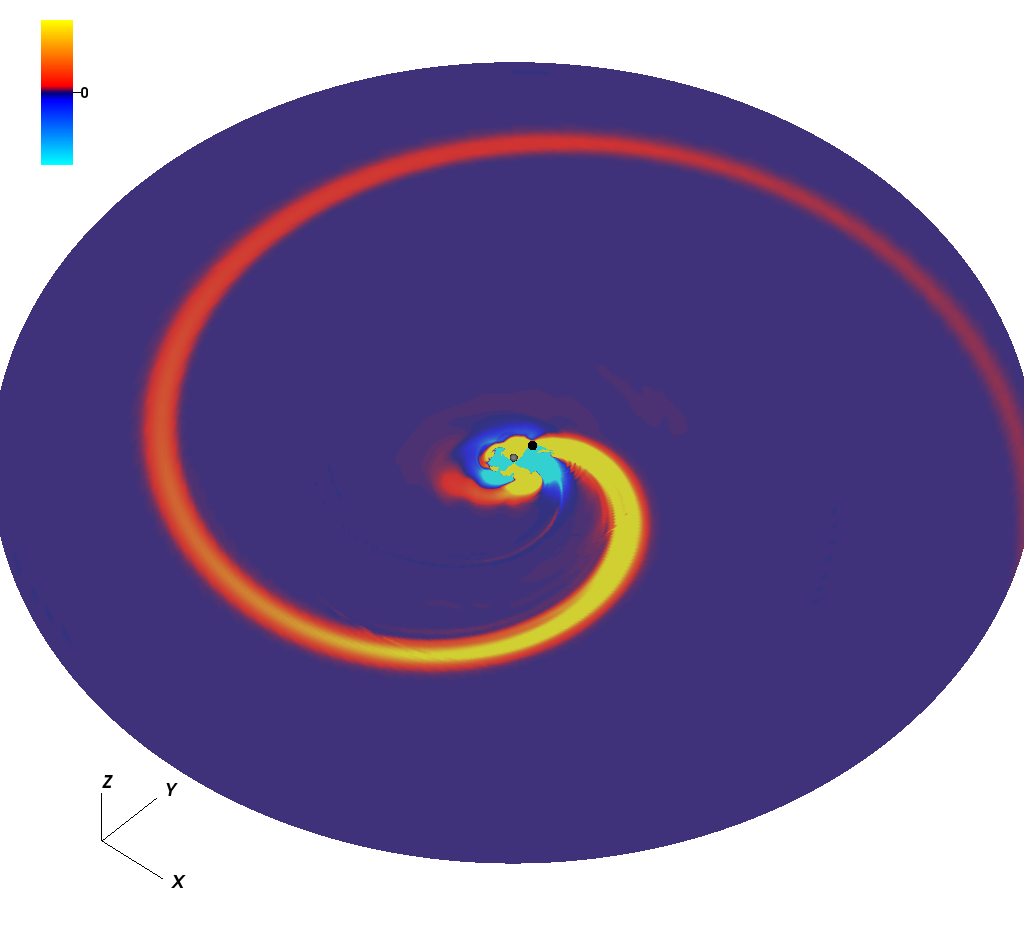}
\\
\vspace{-1cm}
\includegraphics[scale=0.31]{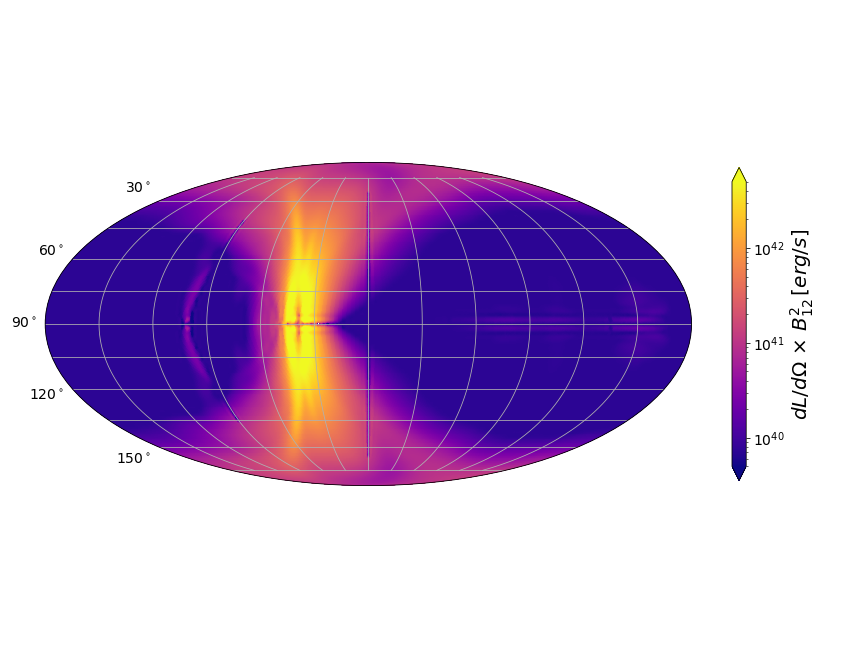}\\
\vspace{-1cm}
\caption{Poynting flux distribution for a BHNS binary in a circular orbit at a separation $r_o = 10 M_{\rm BH}$. Top panel: %3D view of the \MS{It is not 3D view} 
the spiral structure of the flux, shown in color scale at the plane $z=-3M_{\rm BH}$. The gray and black surfaces represent the NS and BH, respectively. Bottom panel: sky-map distribution of Poynting flux at a radius of $r=40 M_{\rm BH}$.
}
 \label{fig:PF}}
\end{figure}

The Poynting flux distribution concentrates along the same spiral arm of the red component on the equatorial CS, as can be seen in Fig.~\ref{fig:PF}. The reason for this is that this region of the magnetic loops is where most of the twisting concentrates, and thus, the strength of the magnetic field, and hence the EM energy, are largely enhanced. 
The bottom panel of Fig.~\ref{fig:PF} presents a skymap distribution of such EM fluxes on a distant sphere around the BH. The intensity peaks on a narrow band %($\sim 15\degree$) 
in the azimuthal direction and within $\sim 60\degree$ from the orbital plane. This would suggest a lighthouse effect being established with the setup orbital frequency, consistent with previous results (e.g., \cite{paschalidis2013, carrasco2020}).
%\FC{we can say something about the degree of anisotropy. In the sense there is still significant luminosity (only approx. one order of magnitude smaller) around the polar regions.}
%We note that significant luminosity can be found around the polar regions as well, only approximately $1$--$2$ orders of magnitude smaller compared to the peak values.

%%%%%%%%%%%%%%%%%%%%%%%%%%%%%%%%%%%%%%%%%%%%%%
%%%%%%%%%%%%%%%%%%%%%%%%%%%%%%%%%%%%%%%%%%%%%%

We compute the luminosity for our late-time numerical solutions by integrating \eqref{eqn:flux} at some distant radius around the BH in the wave-zone (typically, we chose $r = 40 M_{\rm BH}$). We normalize them with the EM luminosity given by the MD radiation formula \cite{ioka2000},
\begin{equation}
 L_{\rm MD} =\frac{4}{15c^5} \mu^2 r_{o}^2 \Omega_{o}^6 ,  \label{eq:zeroth} 
\end{equation}
although it should be taken only as a reference value, since there is no particular reason to chose this scaling in the present GR modeling.

\begin{figure}%[!ht]
\centering{
\includegraphics[scale=0.33]{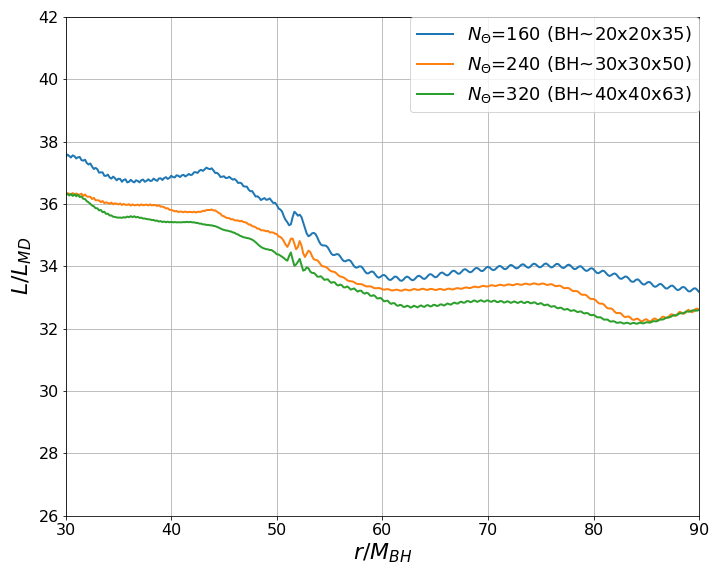}
\caption{EM luminosity for a non-spinning BHNS binary in a circular orbit at an orbital separation $r_o = 10 M_{\rm BH}$. The luminosity is shown as a function of the integration radius for $3$ resolutions, indicated by their values of $N_{\theta}$ (in parenthesis the number of grid-points that cover the BH is indicated). The luminosity is normalized  
by $L_{\rm MD}$ (see Eq.~\eqref{eq:zeroth}). 
}
 \label{fig:convergence}}
\end{figure}

Convergence of the luminosity is assessed by considering the results with three different numerical resolutions. Since CSs extend to a region very close to the BH horizon, one does not expect formal convergence of the numerical solutions in the wave-zone, as they would generally depend on the dissipation treatment given to these CSs. However, by keeping a fixed dissipation rate (controlled by a single parameter in our case), we achieved reasonable numerical convergence for the luminosity. The results are summarized in Fig.~\ref{fig:convergence}, in which the luminosity is shown as a function of the integration radius for the three resolutions. We notice that the lowest resolution with $N_{\theta} = 160$ (the one used throughout this work) is already practically convergent as it differs in less than $5\%$ with respect to the higher resolution runs. 
The luminosity decreases slightly with the increase of the extraction radius, because a part of the Poynting flux is dissipated on the CSs.
%%%%%%%%%%%%%%%%%%%%%%%%%%%%%%%%%%%%%%%%%%%%%%
%%%%%%%%%%%%%%%%%%%%%%%%%%%%%%%%%%%%%%%%%%%%%%

\begin{figure}%[!ht]
\centering{
\includegraphics[scale=0.33]{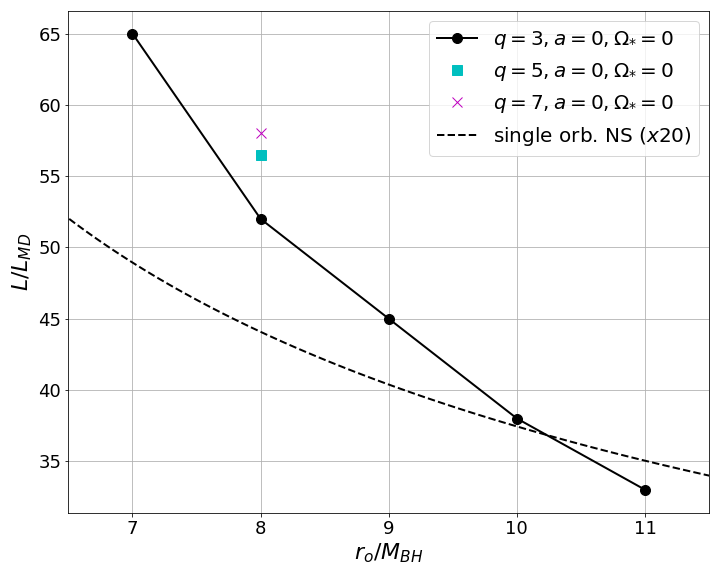}
\caption{EM luminosity for a non-spinning BHNS binary of mass-ratio $q=3$ in a circular orbit at different orbital separations. The results with other mass ratios $q=\{5,7\}$ are also shown (at a given separation, $r_{o} = 8 M_{\rm BH}$) for a comparison. 
The luminosity is computed at $r = 40 M_{\rm BH}$ and normalized by $L_{\rm MD}$ (see Eq.~\eqref{eq:zeroth}). 
A reference scaling (dotted lines) from 
previous results on a single orbiting NS  (i.e., without the BH curvature) \cite{carrasco2020}, re-scaled here by a factor of $20$, is included.
}
\label{fig:Lsepar}}
\end{figure}

Figure \ref{fig:Lsepar} shows the luminosity as a function of the orbital separation.
A reference scaling is presented as well, in order to compare with the results found in the absence of the BH in \cite{carrasco2020}. First, we notice that the luminosity found here is $15$ to $30$ times larger than those from the single orbiting NS scenario, and that the scaling is here steeper towards the closest binary orbits. This indicates that the new effect contributed by the BH curvature is dominant, and thus, the scaling associated with it certainly differs from the reference, Eq.~\eqref{eq:zeroth} (even if accounting for the relativistic corrections found in \cite{carrasco2020}).  
It is important to remark that the luminosity predicted in the unipolar induction (UI) model posses the same scaling as $L_{\rm MD}$ for a non-spinning BH, and only differs by a factor $\sim 5.6$ (larger) for a binary of mass-ratio $q=3$. Thus, as also found in \cite{east2021}, the resulting luminosity is significantly higher than the theoretical expectations from the UI mechanism, especially near the last stable orbits. 

An estimated electric dipole moment in the near zone (approximated as $\int_{r<c/\Omega_o} \mathcal{Q} \, x^i \, d^3 x$, with respect to the CoM) yields higher values by a factor of $5$--$10$ than the magnetic dipole moment of the NS. This indicates that the induced charges and currents within the magnetosphere are likely to be the origin of the high EM luminosity obtained.
As magnetic reconnections occur near the BH, the reconnection site is closer to the NS than in the scenario without BH studied in \cite{carrasco2020} where the CS begin at $\sim c/\Omega_o$. This implies that the magnetic field is here much stronger, and may explain the enhanced luminosity. 

We also note the normalized luminosity depends only very weakly on the mass-ratio, $q$, 
%$q:= M_{\rm BH} / M_{\rm NS}$, 
when re-scaling the problem by the BH mass (at a fixed value of $M_{\rm NS}$). Therefore, these results are quite generic in that respect, and can be extended directly to binaries with higher mass-ratios.

As shown by Eq.~\eqref{eq:zeroth}, $L_{\rm MD}$ is proportional to $r_o^{-7}$ for $a=0$. 
Figure~\ref{fig:Lsepar} suggests that for close orbits, $L$ increases 
as $r_o^{-7-n}$ with $n\agt 1$, and thus, at an innermost orbit 
of $r_o \sim 6M_{\rm BH}$, $L$ is likely to be $\sim 70$--$80L_{\rm MD}$ 
irrespective of the mass-ratio $q$. For a plausible value of $\mu=10^{30}\,{\rm G\,cm^3}$ 
and for plausible masses of the BH and NS, 
$M_{\rm BH}=7.0M_\odot$ and $M_{\rm NS}=1.4M_\odot$, with $r_o=6M_{\rm BH}\approx 74$\,km, $L_{\rm MD}$ is written as
\begin{eqnarray}
 L_{\rm MD} &\approx& 1.3 \times 10^{40}\,{\rm erg/s}
 \left({\mu \over 10^{30}\,{\rm G\,cm^3}}\right)^2 \nonumber \\
 && \times 
 \left({M_{\rm BH} + M_{\rm NS} \over 8.4 M_\odot}\right)^3
 \left({r_o \over 74\,{\rm km}}\right)^{-7},  \label{eq:zeroth1} 
\end{eqnarray}
Thus, near the innermost stable circular orbits, the luminosity $L$ is likely to be enhanced up to $\sim 10^{42}\,{\rm erg/s}$ for the non-spinning BHs. For rapidly spinning BHs with the spin axis aligned with the orbital angular momentum, the smaller orbital radius can be taken. 
The steep dependence of $L$ on $r_o$ indicates that the luminosity
can exceed $10^{44}\,{\rm erg/s}$ if the stable orbits of $r_o = 3M_{\rm BH}$ are allowed (see below, Sec.~\ref{sec:spin}).

%%%%%%%%%%%%%%%%%%%%%%%%%%%%%%%%%%%%%%%%%%%%%%
%%%%%%%%%%%%%%%%%%%%%%%%%%%%%%%%%%%%%%%%%%%%%%

\begin{figure*}%[!ht]
\centering{
\includegraphics[scale=0.22]{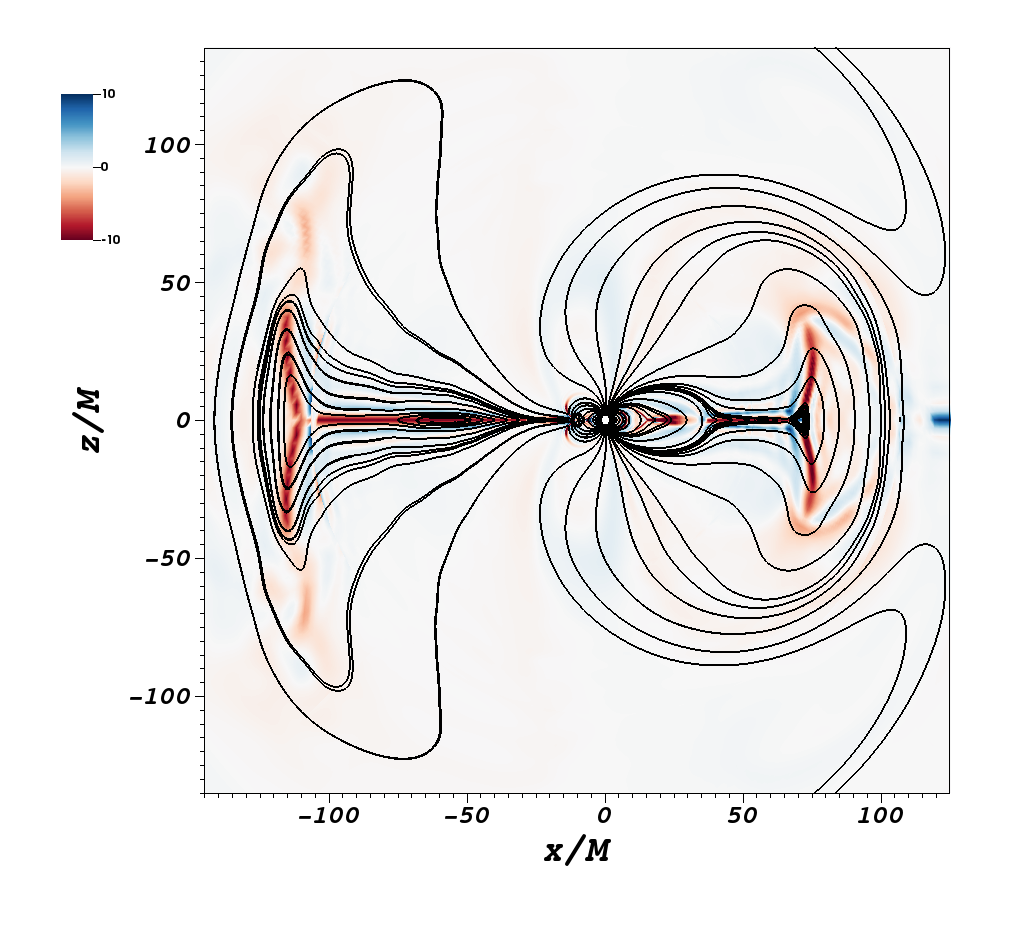} 
\includegraphics[scale=0.22]{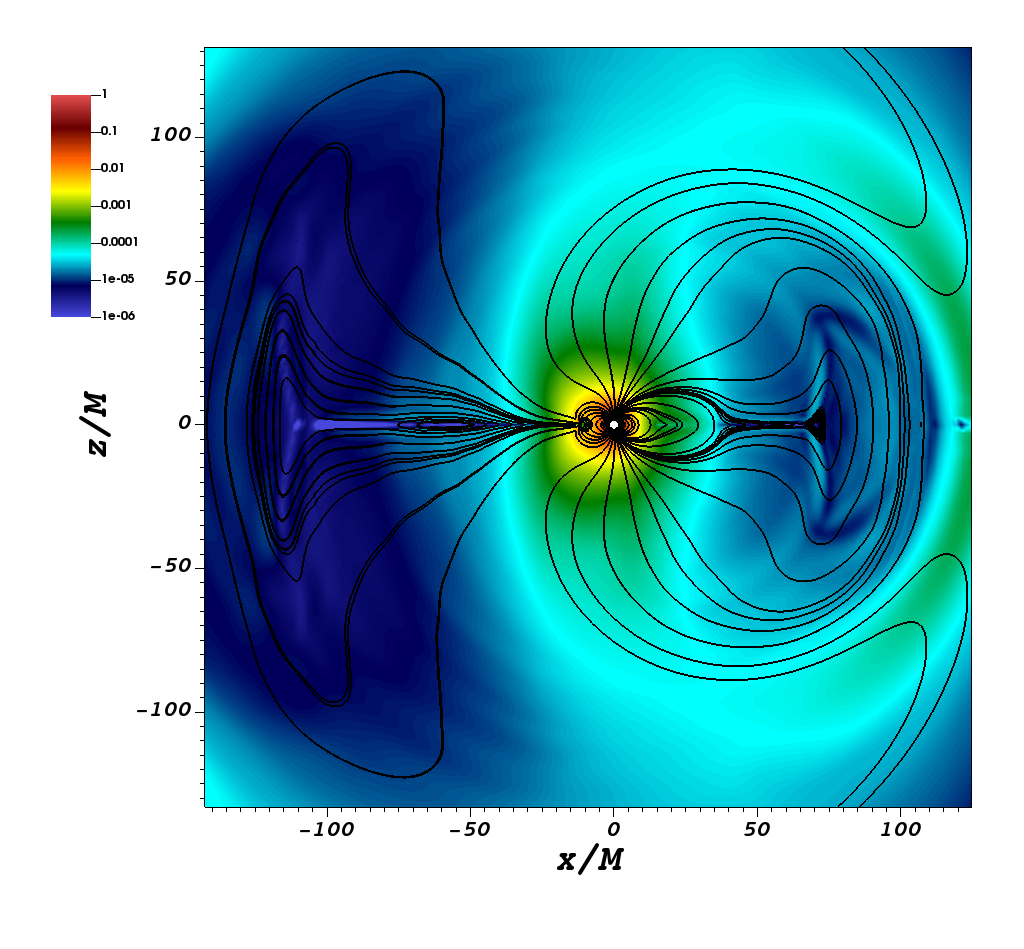}
\\
\includegraphics[scale=0.22]{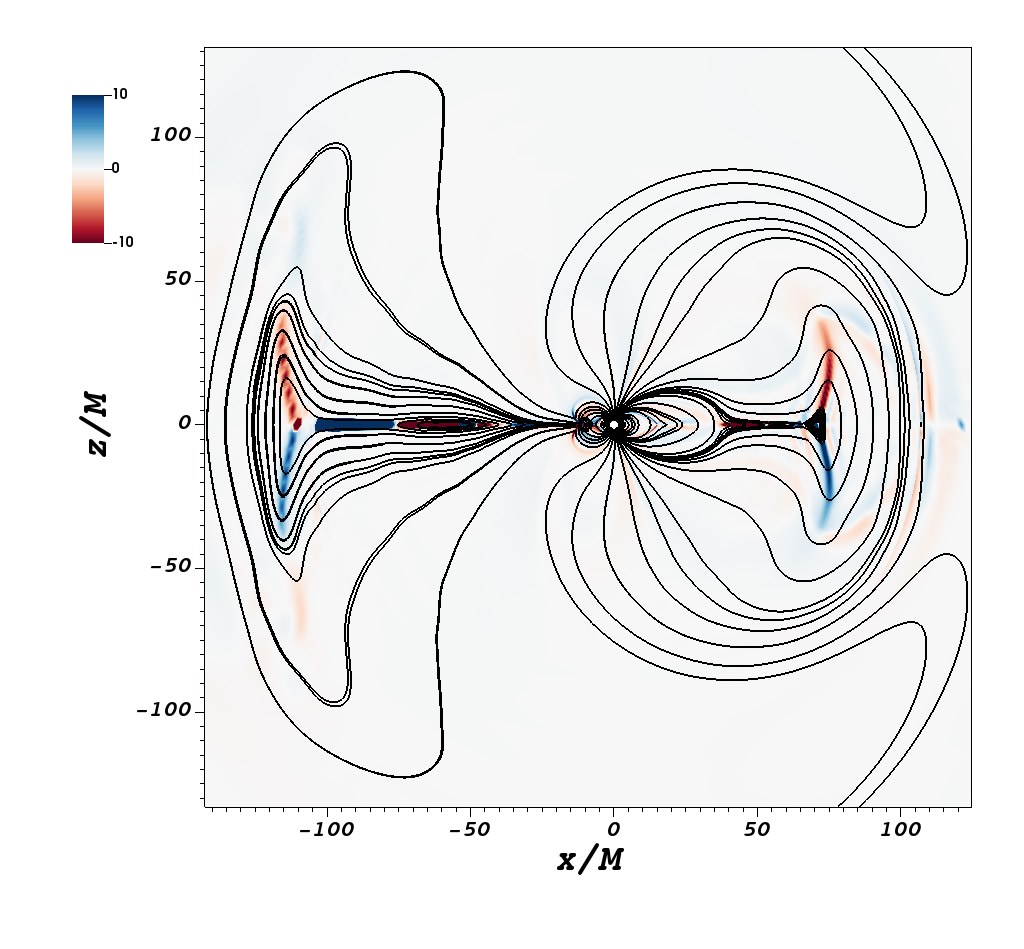} 
\includegraphics[scale=0.22]{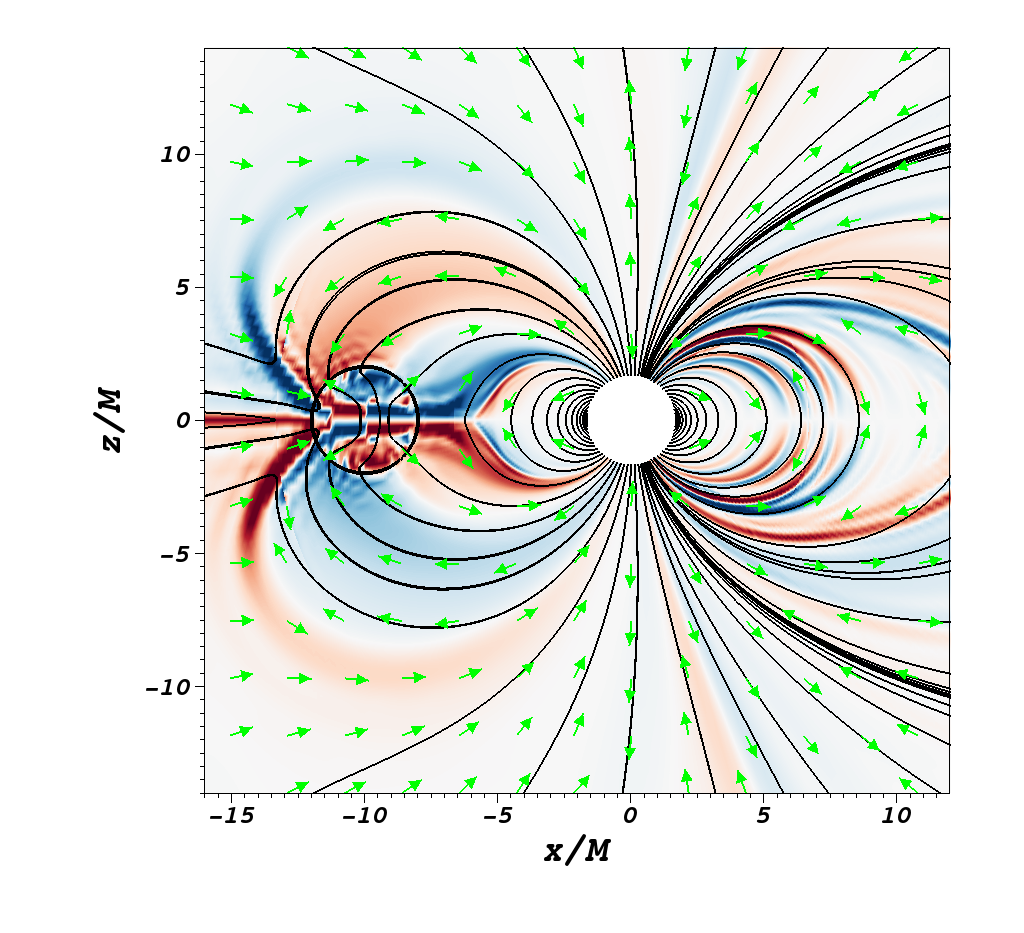}
\caption{Electric charge and current on the co-plane ($x$-$z$ plane) for a BHNS binary in a circular orbit at an orbital separation $r_o = 10 M_{\rm BH}$. 
The numerical solution on the co-plane after $2$ orbits is displayed.
Charge density distribution (top-left panel) and parallel electric currents (bottom-left panel) are normalized by $\Omega_o B / 2\pi c$ and $\Omega_o B / 2\pi$, respectively.
%Electric currents along magnetic field lines are shown with the normalization by $\Omega_o B / 2\pi$.
Top-right plot shows the magnetic field strength, normalized by its value at the poles 
while bottom-right panel is a zoom-in version of the bottom-left one.
The circle with its center at $x=z=0$ shows the NS surface, and the thick black circle (located at $x <0$) shows the BH horizon;
The green arrows in the bottom-right are plotted to emphasize the direction of the currents, which are suggestive of the electronic circuit predicted by the UI model. 
The open circle around the origin shows the NS and the circle of its center at $x=-10M_{\rm BH}$ is the BH horizon. 
}
 \label{fig:co-plane}}
\end{figure*}

We now continue our detailed description of the magnetosphere, by observing the late time numerical solutions at a particular plane intersecting both compact objects. Figure \ref{fig:co-plane} shows charge density, magnetic field strength, and parallel electric currents (Eq.~\eqref{eq:Jparallel}), together with some magnetic field lines projected onto this co-plane. 
The CSs are clearly found in the images, both the spiral CS on the orbital plane and the out-of-the-plane component (seen at the left side of the BH). The figure also presents a clean view of the plasmoids (i.e., their projection onto the co-plane): see, for instance, all the closed magnetic loops to the left of the BH in the left panel. 
%
%%%%%%%%%%%%%%%%%%%%%%%%%%%%%%%%%%%%%%%%%%%%%%
%%%%%%%%%%%%%%%%%%%%%%%%%%%%%%%%%%%%%%%%%%%%%%
%
As seen before, both charge and current densities are very intense (far beyond the GJ values) along the CSs. There is also significant current flowing along those field lines that connect the BH to the NS. Such ``electric circuit'', depicted more clearly on the right panel of Fig.~\ref{fig:co-plane}, is highly reminiscent of the UI mechanism (e.g., \cite{hansen2001,lyutikov2011electro,lai2012dc,piro2012}). 

One may then ask whether the simplified description of the UI model is able to capture the main dynamics and to explain the luminosity which we measure in the wave zone.
The main question is perhaps how the energy provided by the DC circuit could be transported to large distances, since Alfv{\'e}n waves would --in principle-- remain trapped within the flux-tube connecting the two compact objects. In this respect, the magnetic reconnections which we observe in the vicinity of the BH (not taken into account in the UI model) may play a fundamental role in allowing EM energy to escape. %, carried by the plasmoids. 
Such extreme bending of magnetic field lines near the BH equator has been noticed before in the context of boosted BHs moving across a magnetized plasma \cite{palenzuela2010dual, Luis2011}, and identified as a pure gravitational effect in \cite{Boost}. Here, this phenomenon could circumvent the theoretical expectation that asserts the azimuthal twist of the flux-tube in the DC circuit for a BHNS binary is $\lesssim 1$ \cite{lai2012dc}.
%Moreover, as discussed before, the luminosity is higher than the predicted by UI.

Nevertheless, we do find an approximate realization of the UI mechanism in these strong currents flowing inside the flux-tube, which can be regarded as an additional contribution to the energy budget (i.e., complementary to the luminosity measured at larger distances). This contribution is expected to propitiate the generation of gamma-rays from synchrotron/curvature radiation, as well as thermal X-ray emissions from the accelerated particles that hit the NS surface forming hot-spots (see, e.g., \cite{mcwilliams2011}).

%%%%%%%%%%%%%%%%%%%%%%%%%%%%%%%%%%%%%%%%%%%%%%
\begin{figure}%[!ht]
\centering{
\includegraphics[scale=0.15]{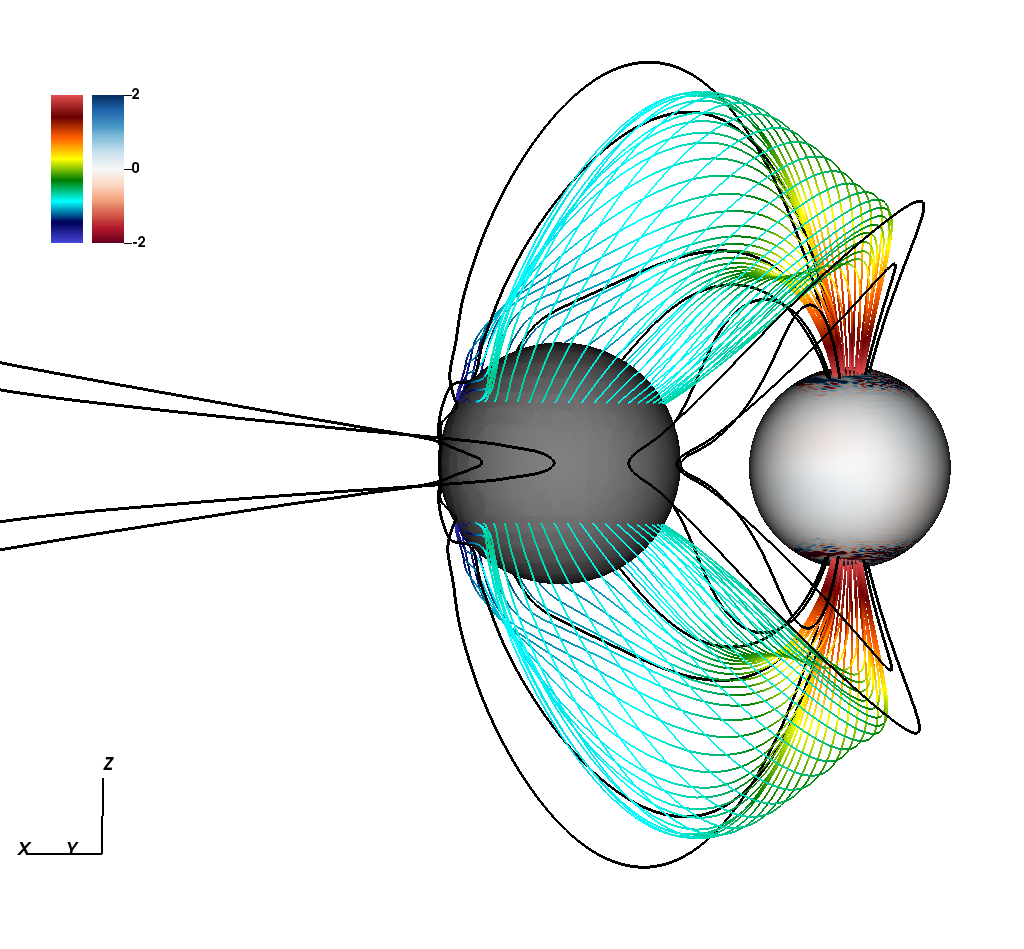}
\\
\includegraphics[scale=0.15]{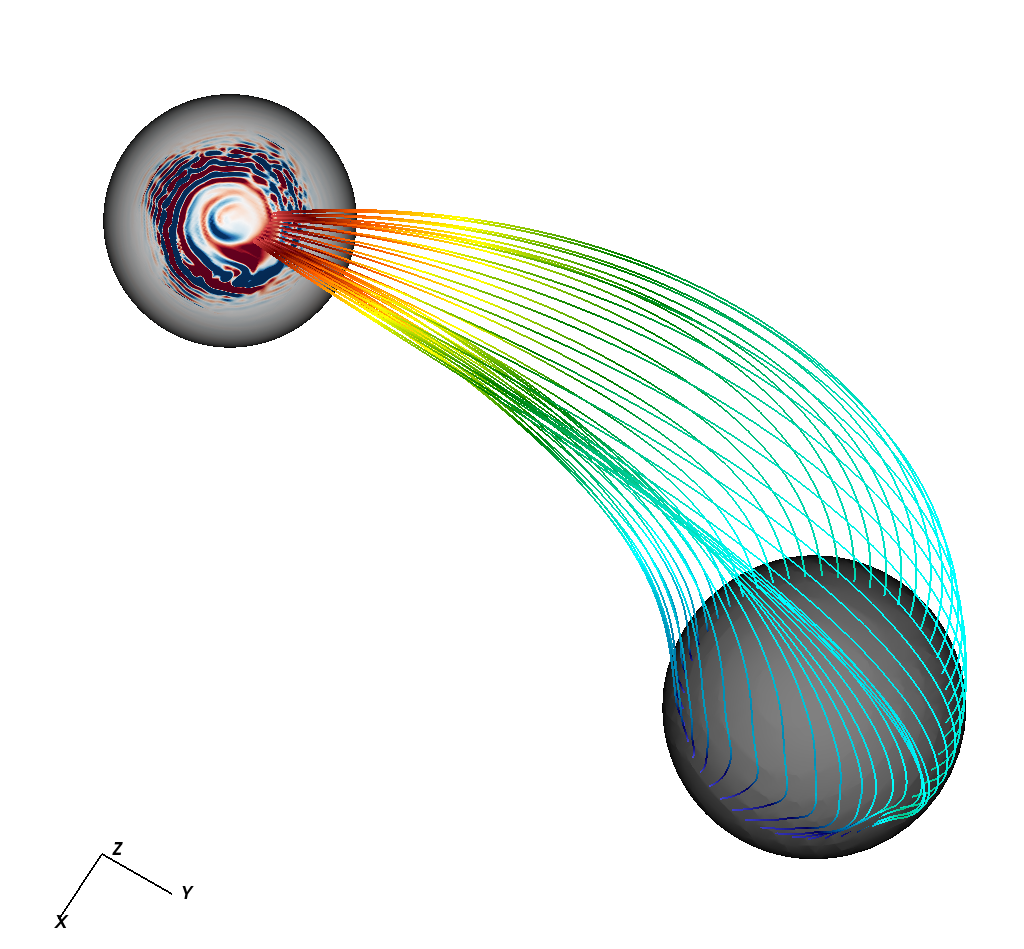} 
\caption{Hot-spots induced on a BHNS binary in a circular orbit 
at an orbital separation $r_o = 10 M_{\rm BH}$. Plotted are 
representative magnetic field lines, some threading the BH horizon with color indicating their strength (in logarithmic scale) and some nearby field lines shown in black. The black sphere represents the BH horizon, while the gray sphere does the NS surface. 
The color map on the NS surface in the bottom panel illustrates the intensity of the magnetospheric currents reaching the NS.
}
 \label{fig:PCs}}
\end{figure}
%%%%%%%%%%%%%%%%%%%%%%%%%%%%%%%%%%%%%%%%%%%%%%
%
Figure~\ref{fig:PCs} presents a closer look at the main flux-tube by showing some representative magnetic field lines, along with the parallel currents reaching the surface of the NS. %It illustrates how the interaction with the BH curvature produce strong currents flowing within the bundle that connect the two compact objects. 
We can identify a polar cap region enclosing these currents at the stellar surface, showing turbulent-like features due to the impact of magnetic reconnections (especially those taking place near the BH). There are two spots at each hemisphere, linked to the ingoing/outgoing currents of the 'instantaneous' electric circuit.
We here estimate the local temperature of such hot-spots as expected from bombardment of relativistic particles, following a simple model from \cite{lockhart2019x} in which the kinetic energy deposition is estimated by assuming that a fixed fraction $ \kappa $ of the total current density $|\vec{J}|$ is carried by relativistic particles with averaged Lorentz factor $\bar{\gamma}$ traveling towards the stellar surface (see also \cite{baubock2019} for details). The rate at which this energy is deposited is finally equated to the power radiated as a black body, yielding:
\begin{equation}
 T \simeq \left( \frac{c^2 m_e \kappa (\bar{\gamma}-1)}{e \sigma} \right)^{1/4} |\vec{J}|^{1/4}   ,
\end{equation}
where $m_e$, $\sigma$, and $e$ denote the electron mass, Stefan-Boltzmann constant and elementary charge, respectively. 

Plugging the numbers from our simulations, we get
\begin{equation}
{\displaystyle T \approx 0.5 \, {\rm keV} 
\left[ \frac{\kappa \bar{\gamma}}{10^4} \right]^{1/4} 
\left[ \frac{B_{\rm pole}}{10^{12}\,{\rm G}} \right]^{1/4} 
\left[ \frac{\Omega_o}{10^{3}\,{\rm rad/s}} \right]^{1/4}} 
\end{equation}  
for a binary of mass-ratio $q=3$ at $r_o=10M_{\rm BH}$ assuming that the 
current density is $20$ times larger than the GJ value. We note that 
the temperature could be higher if the electron-positron pair creation occurs 
near the polar region efficiently. 
%\FC{\cite{mcwilliams2011} estimates about one order of magnitude larger spot temperatures $T_{\rm spot}$ (evaluating with our parameters).}

\subsection{Spin effects}\label{sec:spin}

%%%%%%%%%%%%%%%%%%%%%%%%%%%%%%%%%%%%%%%%%%%%%%
\begin{figure*}%[!ht]
\centering{
\includegraphics[scale=0.22]{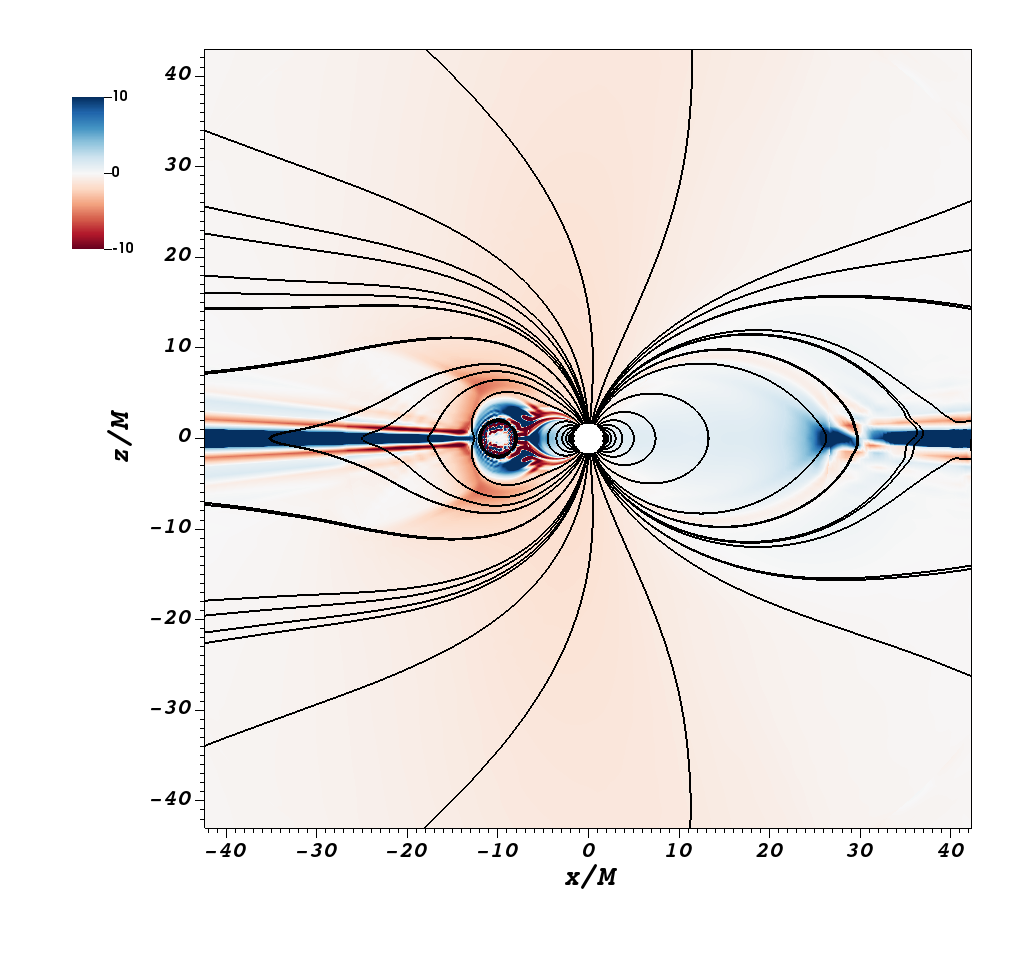} 
\includegraphics[scale=0.22]{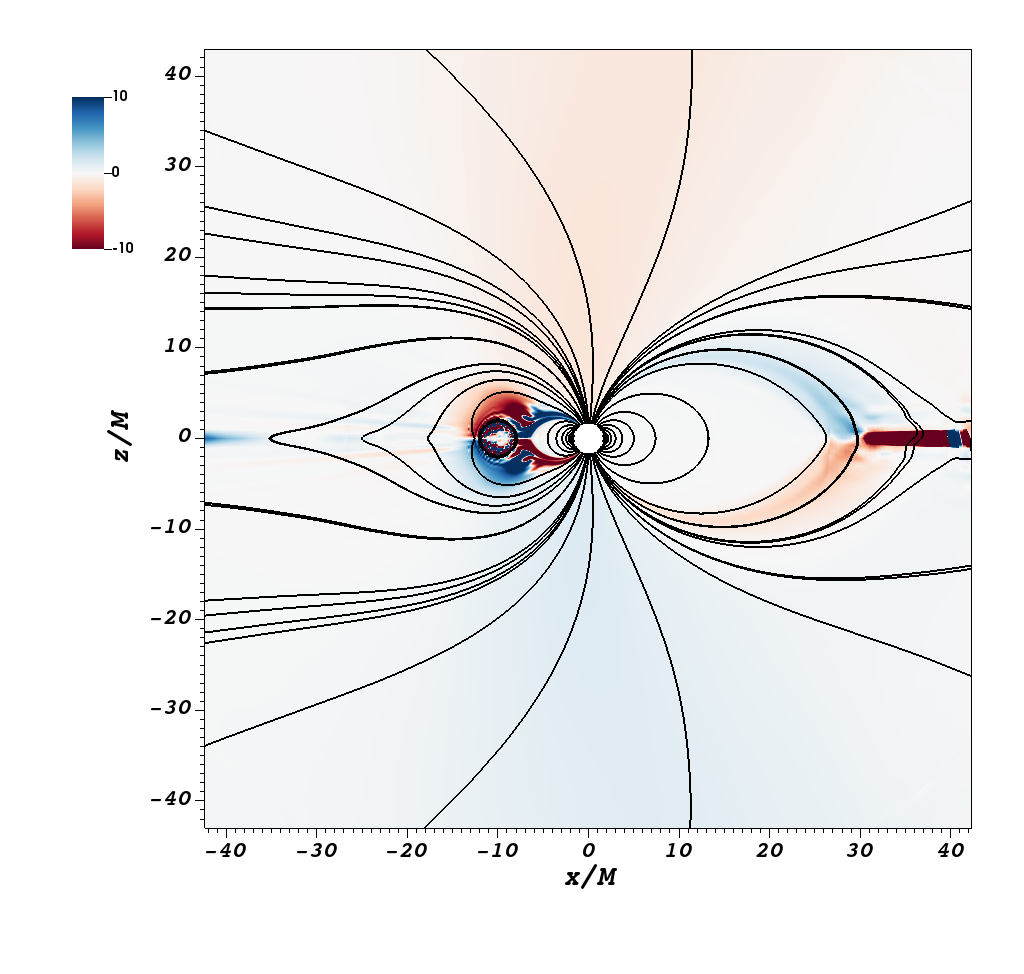}
\\
\hspace{0.5cm}
\includegraphics[scale=0.18]{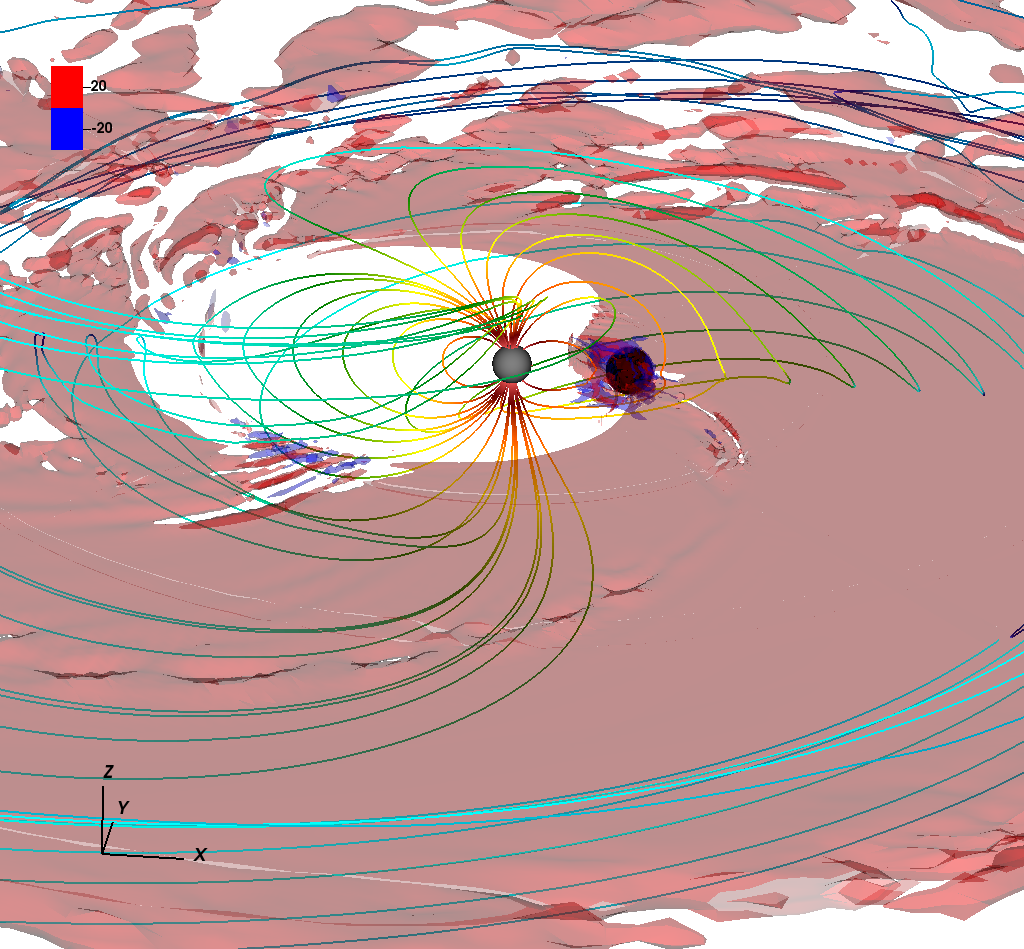}
\hspace{1cm}
\includegraphics[scale=0.18]{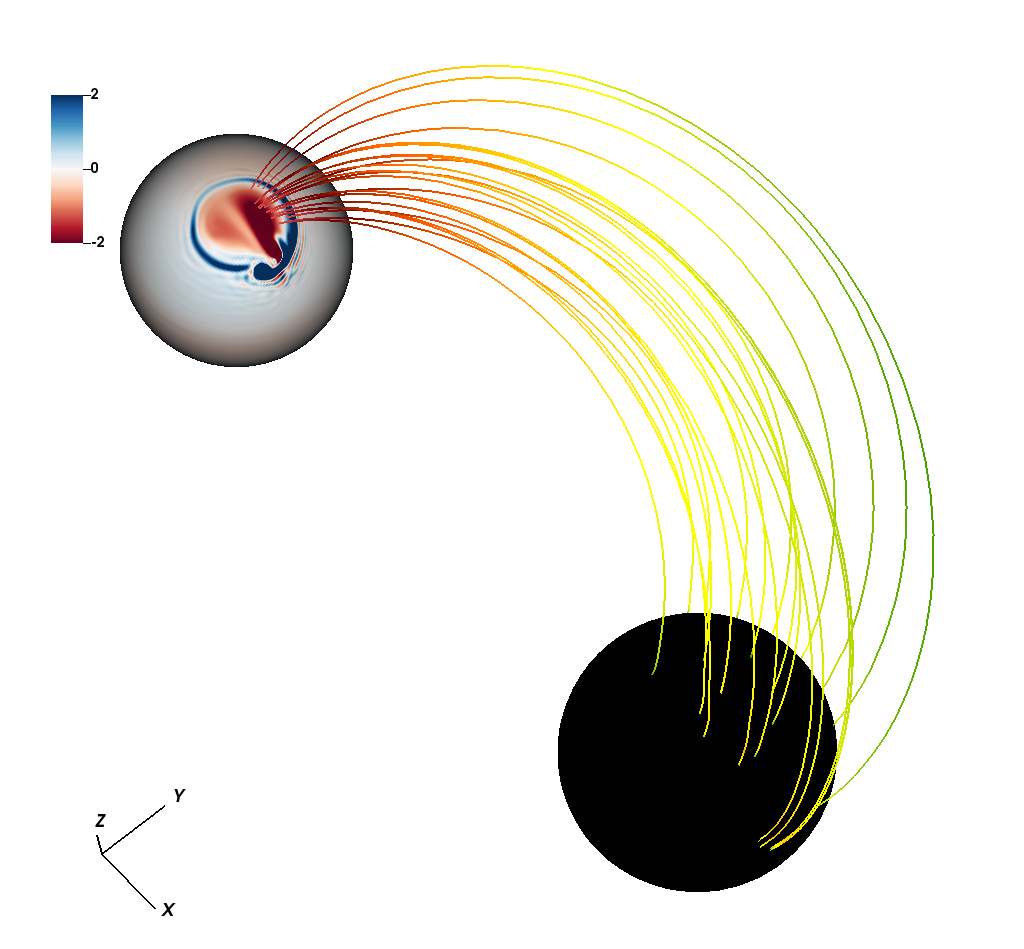}
\caption{
Magnetosphere of a BHNS binary in a circular orbit at an orbital separation $r_o = 10 M_{\rm BH}$ with synchronized NS spin, $\Omega_{*} = \Omega_o$, after $3$ orbits. 
Top panel: Charge density distribution (left) and parallel electric current (right) on the co-plane, normalized by $\Omega_o B / 2\pi c$ and $\Omega_o B / 2\pi$, respectively. The circle with its center at $x=z=0$ shows the NS surface, and the thick black circle (located at $x <0$) shows the BH horizon.
Bottom panel: 3D view of the magnetosphere (left) with contours of the charge density signaling the equatorial CS. Induced polar cap on the stellar surface (right), indicated by a color map of the  magnetospheric currents reaching the NS.
} 
 \label{fig:spin-NS}}
\end{figure*}
%%%%%%%%%%%%%%%%%%%%%%%%%%%%%%%%%%%%%%%%%%%%%%
\begin{figure}%[!ht]
\centering{
\vspace{-1cm}
\includegraphics[scale=0.31]{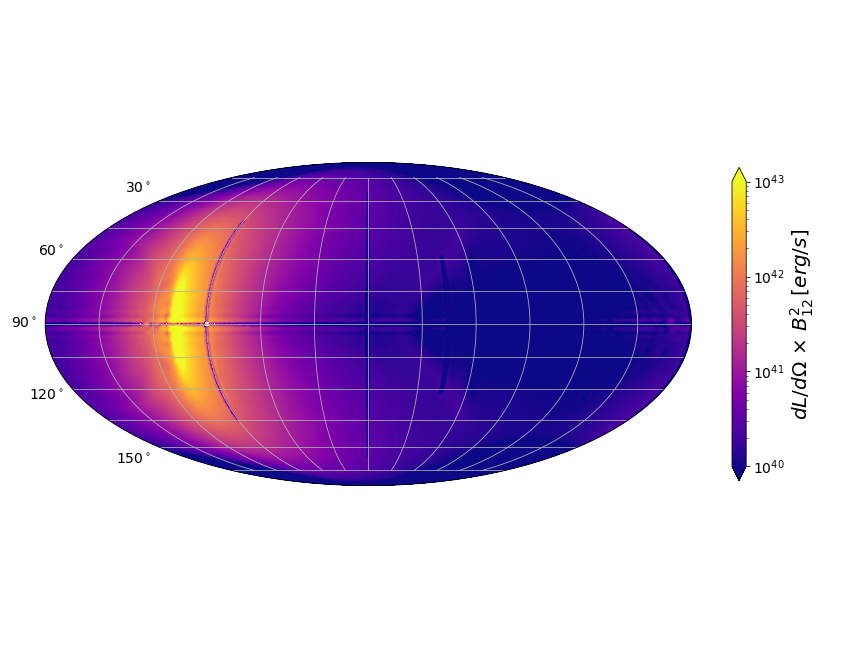}\\
\vspace{-1cm}
\caption{Sky-map distribution of Poynting flux at a radius of $r=40 M_{\rm BH}$ for a BHNS binary in a circular orbit at an orbital separation $r_o = 10 M_{\rm BH}$ with synchronized NS spin, $\Omega_{*} = \Omega_o$. 
}
 \label{fig:NS-spin}}
\end{figure}
%%%%%%%%%%%%%%%%%%%%%%%%%%%%%%%%%%%%%%%%%%%%%%

In this section, we pay attention to the cases in which there is BH or NS spin aligned to the orbital angular momentum. 
We first study the effect of the NS spin, which is expected to be less relevant from the astrophysical perspective, since in the late inspiral phase tidal locking is unlikely to be achieved and the orbital frequency is usually much higher than the NS spin frequency \cite{bildsten1992}.
Here, we only look at the most extreme case in which the spin of the NS is as high as the orbital frequency (i.e., $\Omega_* = \Omega_o$) in a circular orbit at fixed orbital separation $r_o = 10 M_{\rm BH}$. 
For this case, the magnetosphere co-rotates with the orbital motion, and hence, the magnetic field lines penetrate the BH horizon in a stationary manner. This situation is highly different from that considered in the previous section, and thus, the properties of the magnetosphere are also expected to be modified significantly.

We find an EM luminosity of $L \sim 56 L_{\rm MD}$ (at $r=40 M_{\rm BH}$), which is about $50\%$ higher than the one obtained for the same setting with non-spinning NS.
This luminosity is exactly twice the spin-down luminosity of an isolated pulsar of the same period ($\sim 4$\,ms for a binary of mass-ratio $q=3$), estimated as $L_{\rm pulsar} \approx \mu^2 \Omega_{*}^4$~\cite{spitkovsky2006}.
%Thus, we further notice that the luminosity produced by the irrotational binary is comparable (and even higher, as in this case) to the luminosity of an isolated pulsar at the same frequency.
Very similar enhancement with respect to the pulsar spin-down luminosity was already observed in~\cite{carrasco2020} for analogous configuration but without the BH. This suggests that the dynamical role played by the BH is here subdominant in contrast to the case of irrotational BHNS binaries discussed in the previous section. Indeed, the co-rotation of the NS with the orbit diminishes the relative velocity among the plasma and the BH, dramatically reducing its impact on the magnetic-field structure. The magnetosphere, thus, looks overall rather similar to that of an aligned pulsar but with the spiral arm structure in its Poynting flux resulting from the orbital motion. An equatorial CS develops near the BH (similar to the irrotational case) and close to the LC at the opposite side of the NS, with all the magnetic field lines opening-up beyond that point (see Fig.~\ref{fig:spin-NS}). 

There is, however, some additional enhancement of the magnetic field strength for those field lines threading the BH, associated with the current-carrying flux-tube between the two compact objects. The polar cap regions at the NS surface, being dominated in this case by the NS spin, are slightly modified by these flux-tube currents.

A sky-map distribution of Poynting flux is displayed in Fig.~\ref{fig:NS-spin} for such tidally locked BHNS binary system. It looks very similar to that of the irrotational case (i.e., bottom panel of Fig.~\ref{fig:PF}) but with its intensity being even more concentrated towards the orbital plane.\\

In the following, we switch-on the BH spin in the direction perpendicular to the orbit and explore its impact on the magnetosphere.
The dimensionless spin $ a/M_{\rm BH}$ is varied from $-0.99$ to $0.99$ for a circular orbit at fixed orbital separation $r_o = 10 M_{\rm BH}$. Generally speaking, we find that the effect of the orbital motion rather than that of the BH spin determines the magnetosphere and EM luminosity, even for the near-extreme Kerr BHs. In particular, if only the BH spin is present (i.e., without orbital motion) we find a luminosity consistent with the Blandford-Znajek mechanism within the flux-tube connecting both objects; but no Poynting fluxes outside the near-zone. Thus, the orbital motion, with its non-trivial plasma effects close to the BH, provides the mechanism allowing the EM energy to escape to large distances. 

\begin{figure}%[!ht]
\centering{
\includegraphics[scale=0.33]{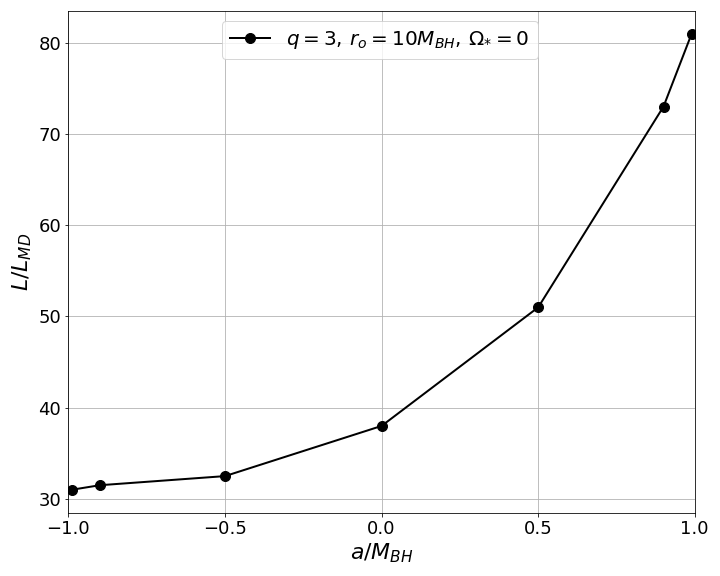}
\caption{EM luminosity for BHNS binaries in circular orbits at separation $r_o = 10 M_{\rm BH}$ for several BH spins. The luminosity is computed at $r = 40 M_{\rm BH}$ and normalized by $L_{\rm MD}$. % (see Eq.~\eqref{eq:zeroth}). 
}
 \label{fig:Lspin}}
\end{figure}

A prograde BH spin --with respect to the orbit-- tends to increase the dynamical effects produced by the orbital motion, while a retrograde spin acts in the opposite way. %by reducing the local velocities of the plasma near the BH. 
It is found, from Fig.~\ref{fig:Lspin}, that the prograde spin is capable of enhancing the luminosity (up to a factor of $\sim2$), whereas a retrogade BH spin slightly (by $\sim 20 \%$) diminishes its value. 
This is again in contrast with the expectations from the UI model (see, e.g., \cite{mcwilliams2011}), in which the trend of $L_{\rm UI}/L_{\rm MD}$ with BH spin is very different from the one found here.

\begin{figure*}%[!ht]
\centering{
\includegraphics[scale=0.15]{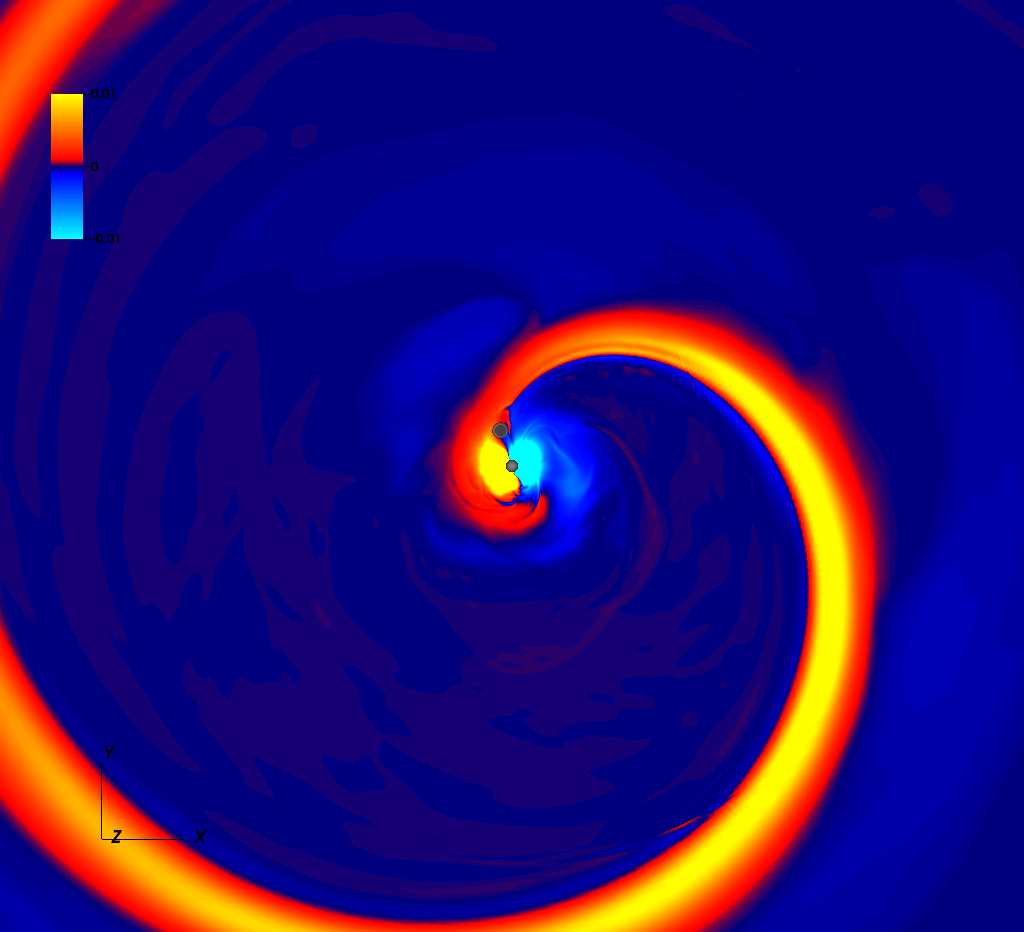} 
\includegraphics[scale=0.15]{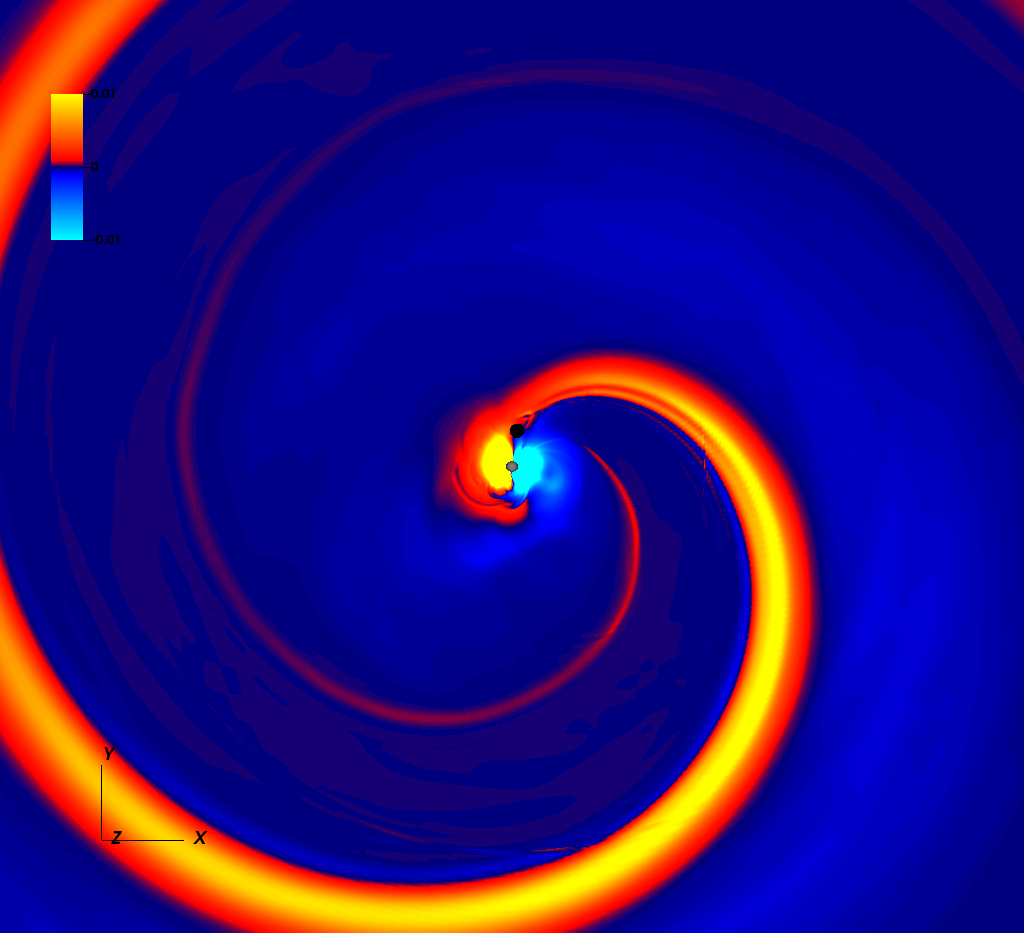}
\includegraphics[scale=0.15]{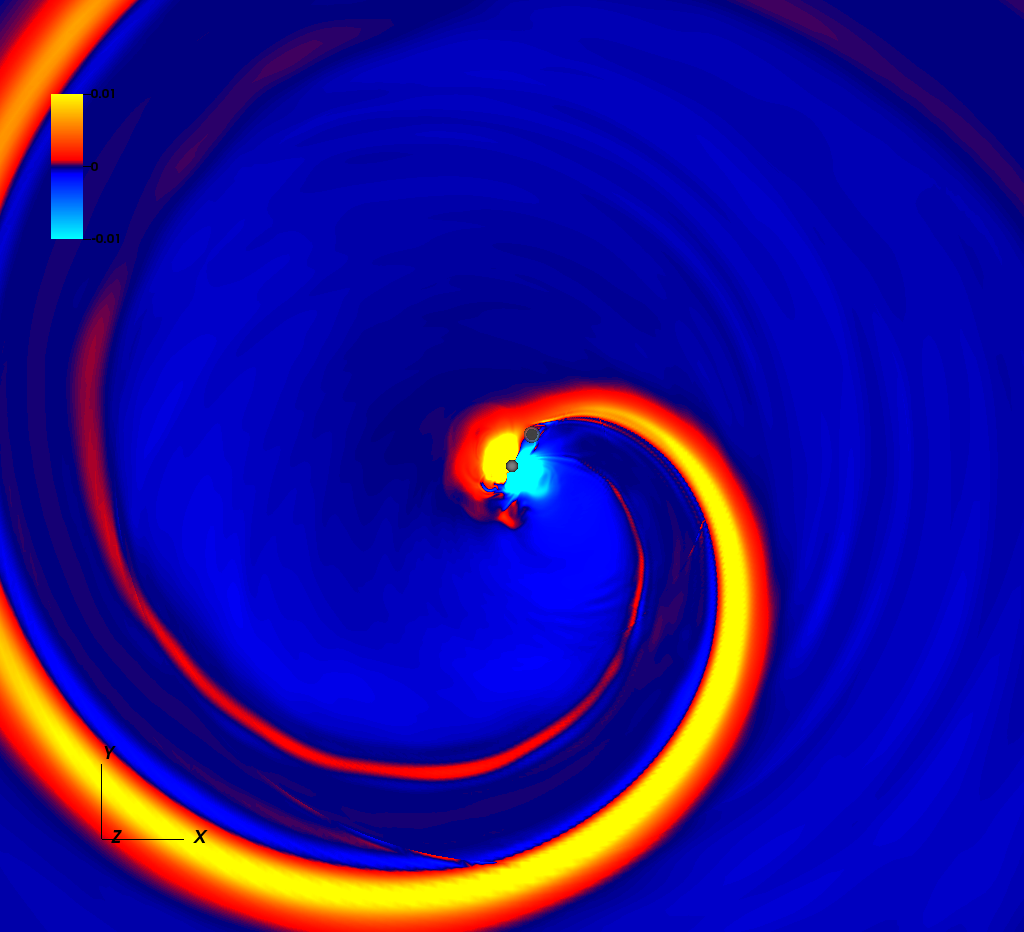}
\\
\includegraphics[scale=0.15]{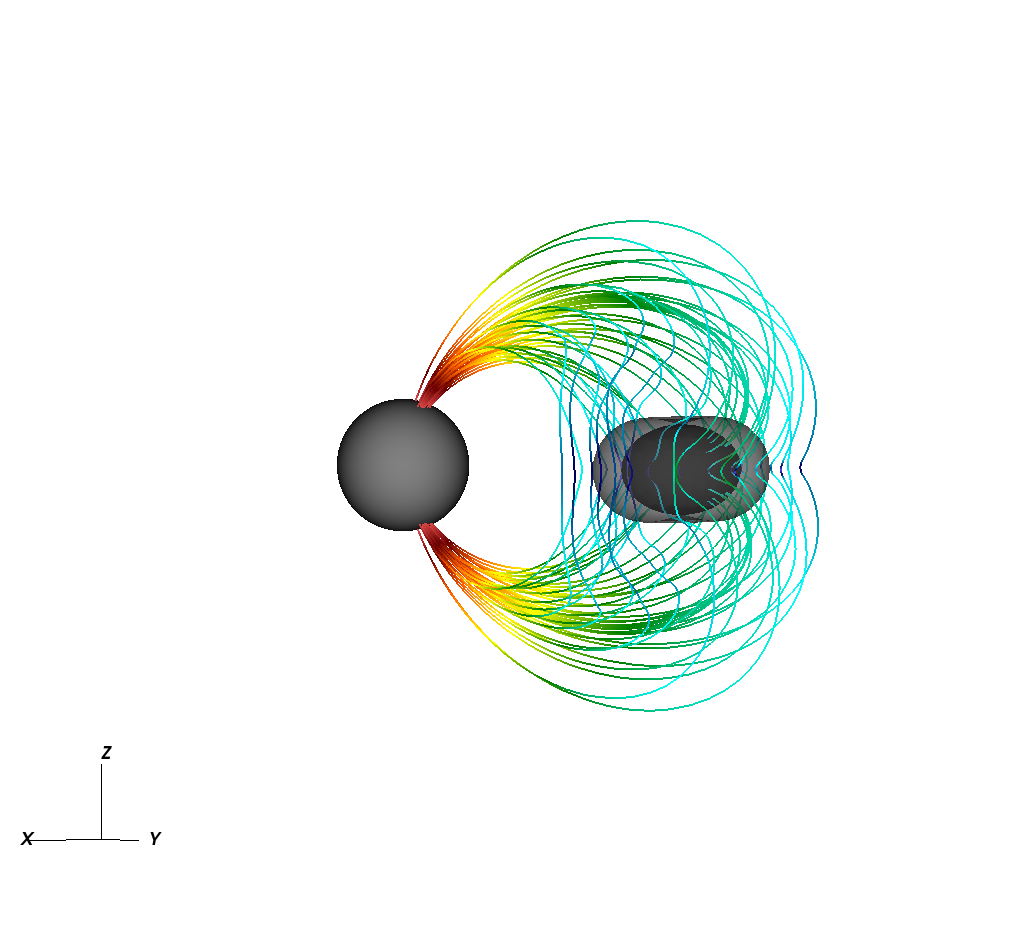} 
\includegraphics[scale=0.15]{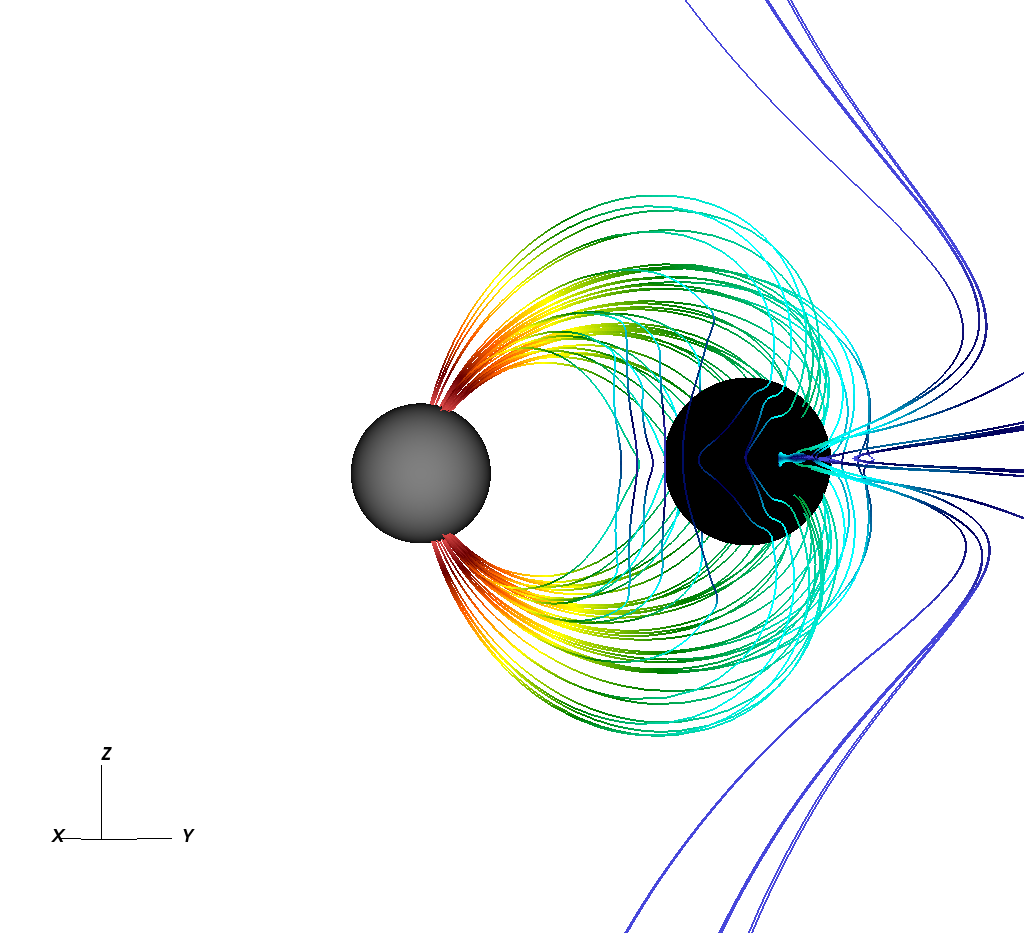}
\includegraphics[scale=0.15]{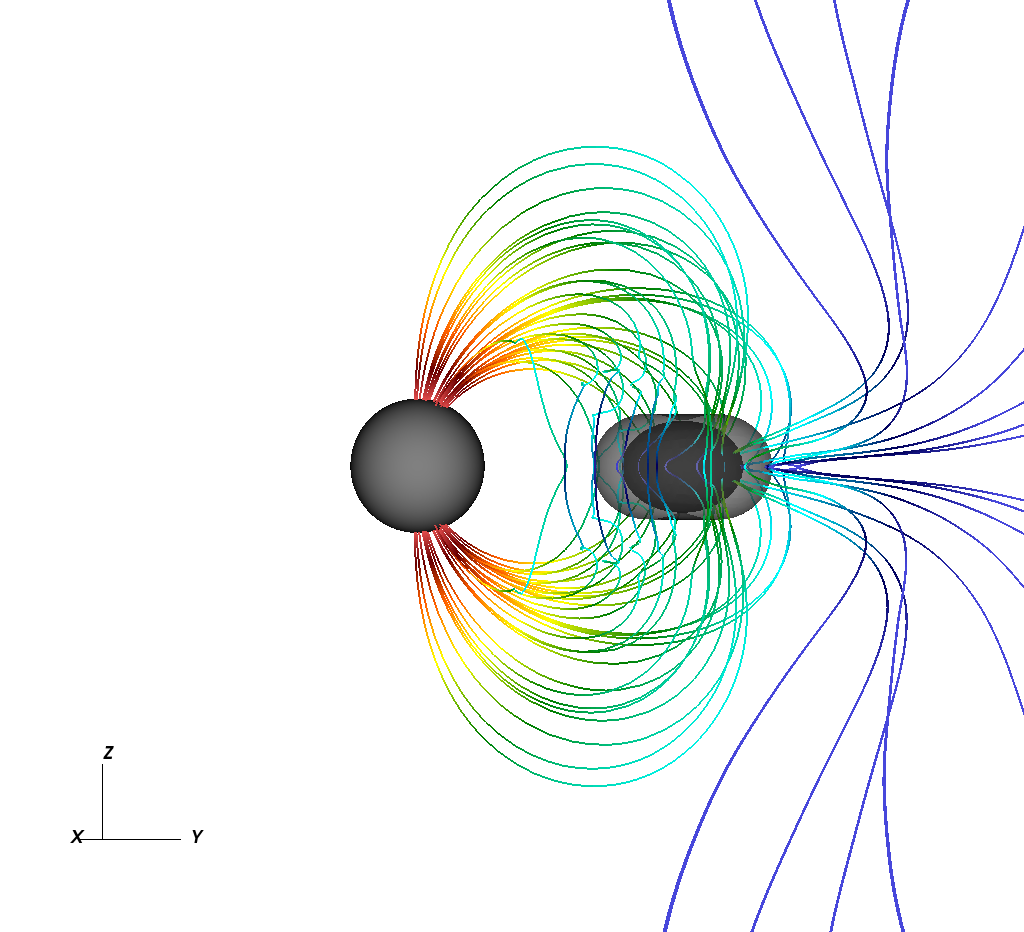}

\caption{Toroidal magnetic field strength for different BH spins: $a_* = -0.99$ (left), $a_* = 0$ (middle) and $a_{*}=0.99$ (right). 
Top panel: the invariant quantity $ F^{*}_{t\phi} \equiv (\partial_{t})^a F^{*}_{ab} (\partial_{\phi})^b $ is displayed on a spatial cut at $z=-3M_{\rm BH}$. Bottom panel: representative magnetic field lines in the vicinity of the BH, with color indicating their strength (in logarithmic scale). The solid gray surface represents the NS, while the BH horizon is indicated in black and the ergoregion by a shaded gray area. 
}
 \label{fig:Bphi}}
\end{figure*}

In order to understand this behavior, we plot in Fig.~\ref{fig:Bphi} the toroidal magnetic field --as defined from the two Killing vectors $\partial_{t}$ and $\partial_{\phi}$ in Kerr spacetime--, $ F^{*}_{t\phi} \equiv (\partial_{t})^a F^{*}_{ab} (\partial_{\phi})^b $. We find that the extra gravitational frame-dragging effect contributes to further twisting of the magnetic field lines near the BH with the prograde spin, leading to stronger toroidal discontinuities that reconnect in a region closer to the BH horizon. On the other hand, the toroidal magnetic field is reduced in the retrograde case and the reconnection site is pushed farther away from the BH, even outside of the ergoregion (see bottom panel of Fig.~\ref{fig:Bphi}). 

For a high prograde spin of $a\agt 0.9M_{\rm BH}$, the 
stable circular orbits are allowed down to $\alt 2.3M_{\rm BH}$. For such orbits, $L_{\rm MD}$ reaches the order of $10^{42}\,{\rm erg/s}$ for plausible values of $\mu$, $M_{\rm BH}$, and $M_{\rm NS}$, as we showed in Eq.~\eqref{eq:zeroth1}.
Thus, the EM luminosity $L$ could exceed $10^{44}\,{\rm erg/s}$ for plausible 
parameters of BHNSs with a high BH spin. For the nearly extreme spin case of $a=0.99M_{\rm BH}$, $L$ is likely to reach the order of $10^{46}\,{\rm erg/s}$ at the innermost stable circular orbit. 

%%%%%%%%%%%%%%%%%%%%%%%%%%%%%%%%%%%%%%%%%%%%%%
%%%%%%%%%%%%%%%%%%%%%%%%%%%%%%%%%%%%%%%%%%%%%%

%%%%%%%%%%%%%%%%%%%%%%%%%%%%%%%%%%%%%%%%%%%%%%
%%%%%%%%%%%%%%%%%%%%%%%%%%%%%%%%%%%%%%%%%%%%%%

\subsection{Misalignment effects}

We now consider situations in which the magnetic moment of the NS is not aligned with the orbital angular momentum, and both compact objects are non-spinning. 
For such irrotational misaligned binaries, the magnetosphere acquires a phase dependency along the orbit, since the position of the BH relative to the inclined dipole of the NS makes an important difference in the structure of the magnetosphere. %the dynamics. %, breaking the previous symmetry of the aligned scenarios. 
%This is clearly reflected in Fig.~\ref{fig:Echi}, where both the EM energy in the domain and the luminosity across a given spherical surface at the wave-zone, present modulations with the orbital frequency. Figure \ref{fig:B1-inclined} illustrates the magnetosphere of a $\chi=15\degree$ misalignment for two instances along the binary trajectory, after $2.75$ and $3$ orbits. In the first one (left panel), the BH is located perpendicular to the misalignment plane of the dipole ($x-z$); whereas in the second plot (right panel), the BH lays at this particular plane. These plots show representative magnetic field lines for each configuration, focusing on the bending effect induced by the strong BH curvature. 
Thus, in contrast to the quasi-stationary aligned configurations, the misaligned solutions are instead quasi-periodical. 

\begin{figure*}%[htp]
	\centering
\includegraphics[scale=0.163]{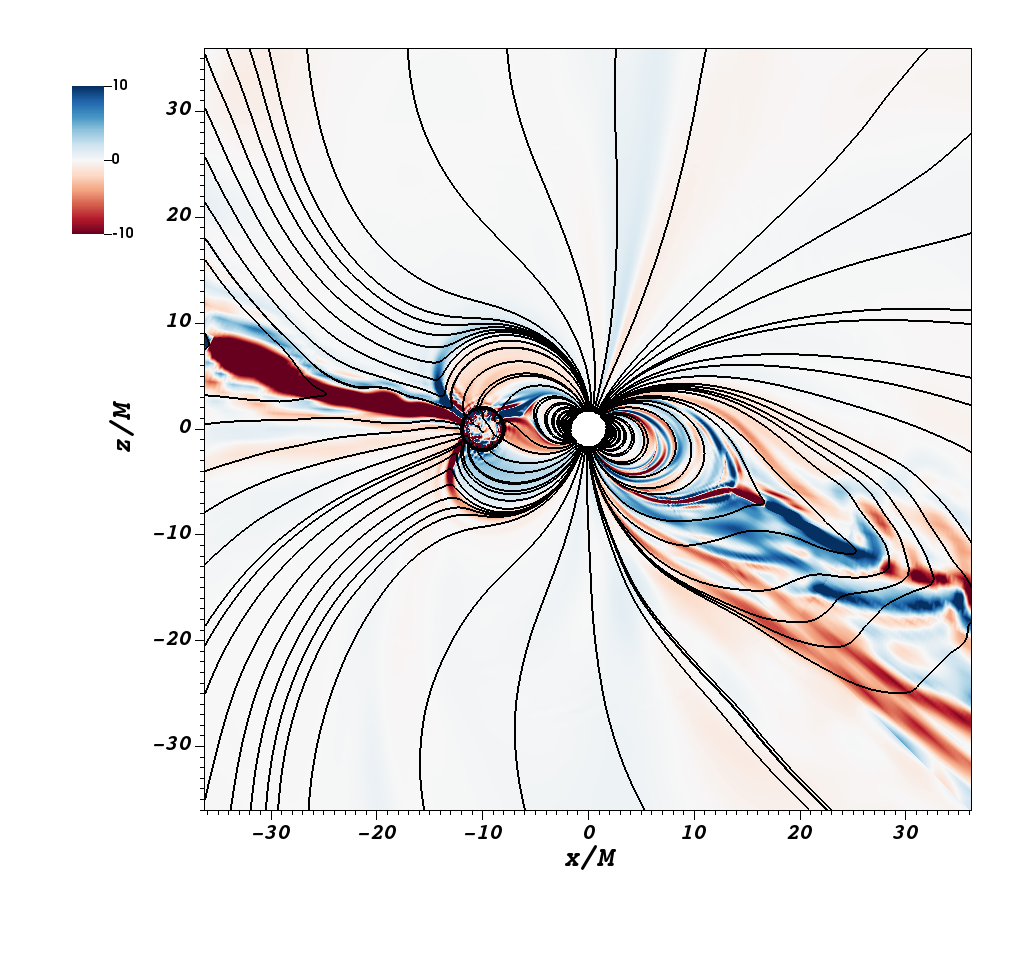}
\includegraphics[scale=0.163]{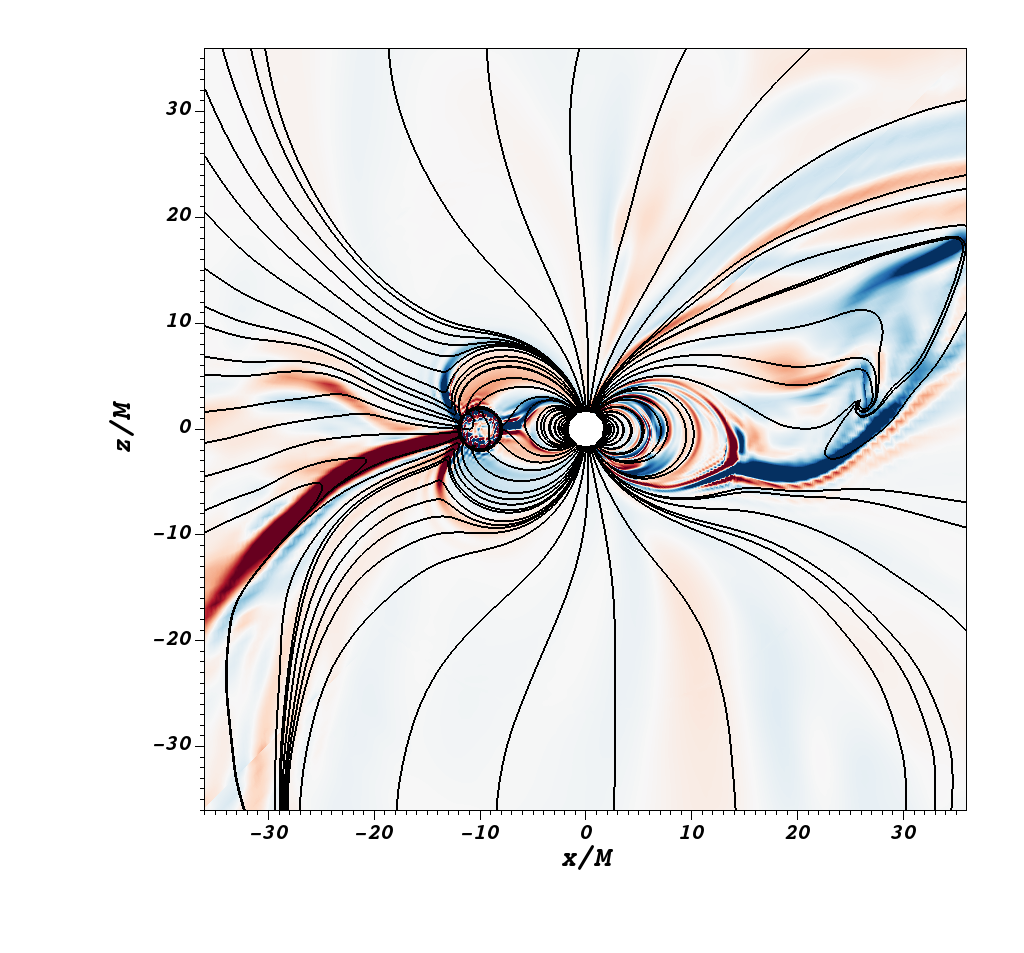}
\includegraphics[scale=0.163]{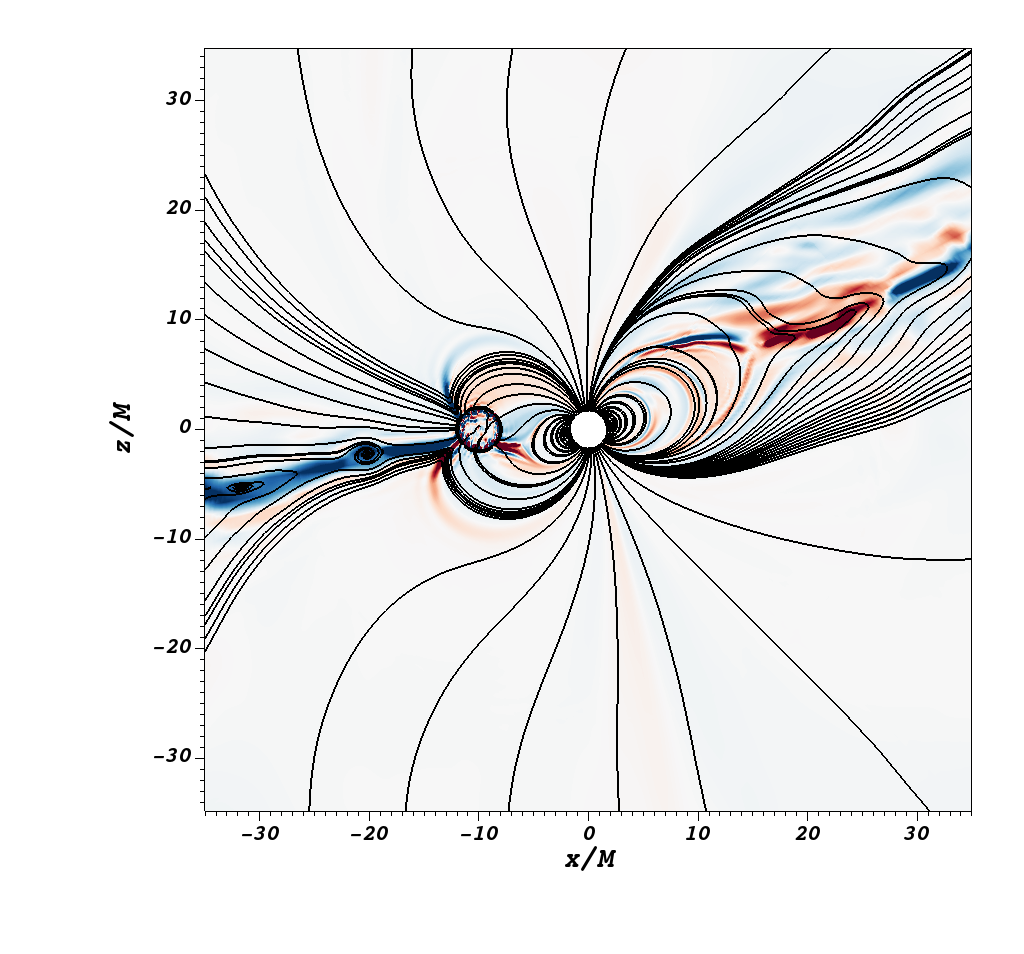}
\\
\includegraphics[scale=0.163]{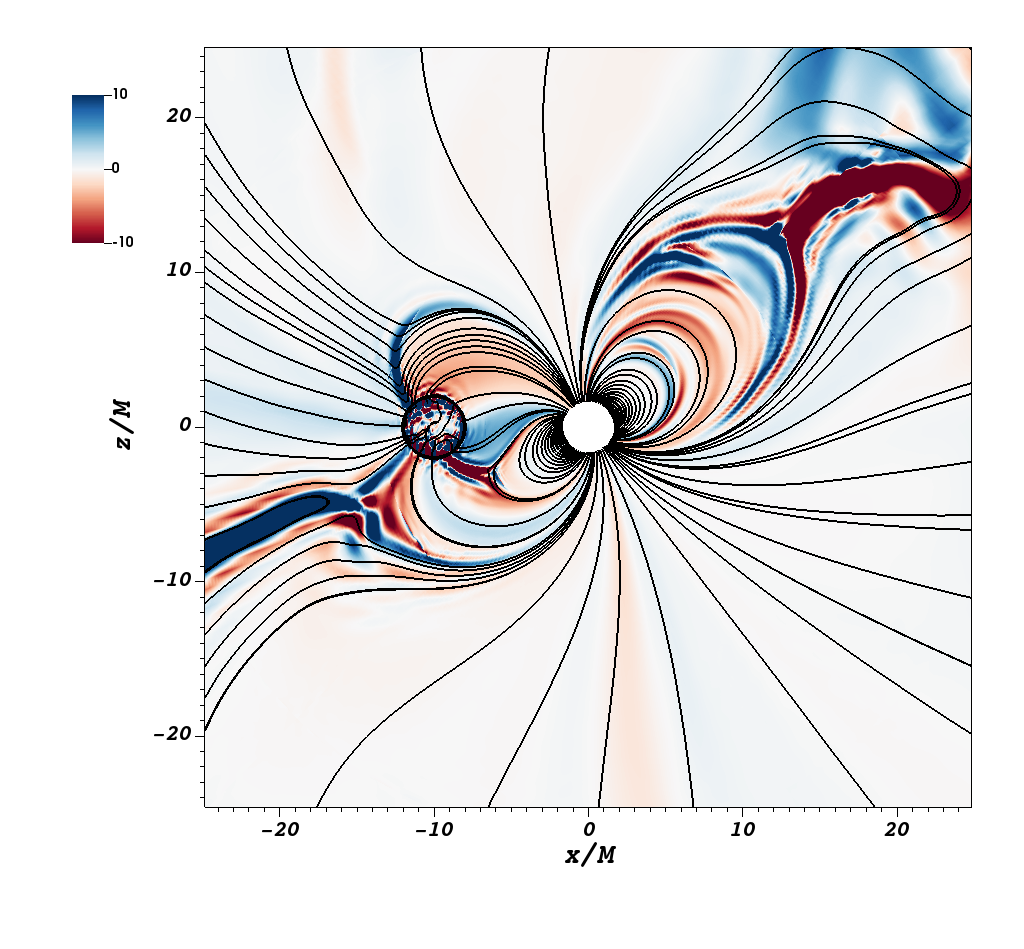}
\includegraphics[scale=0.163]{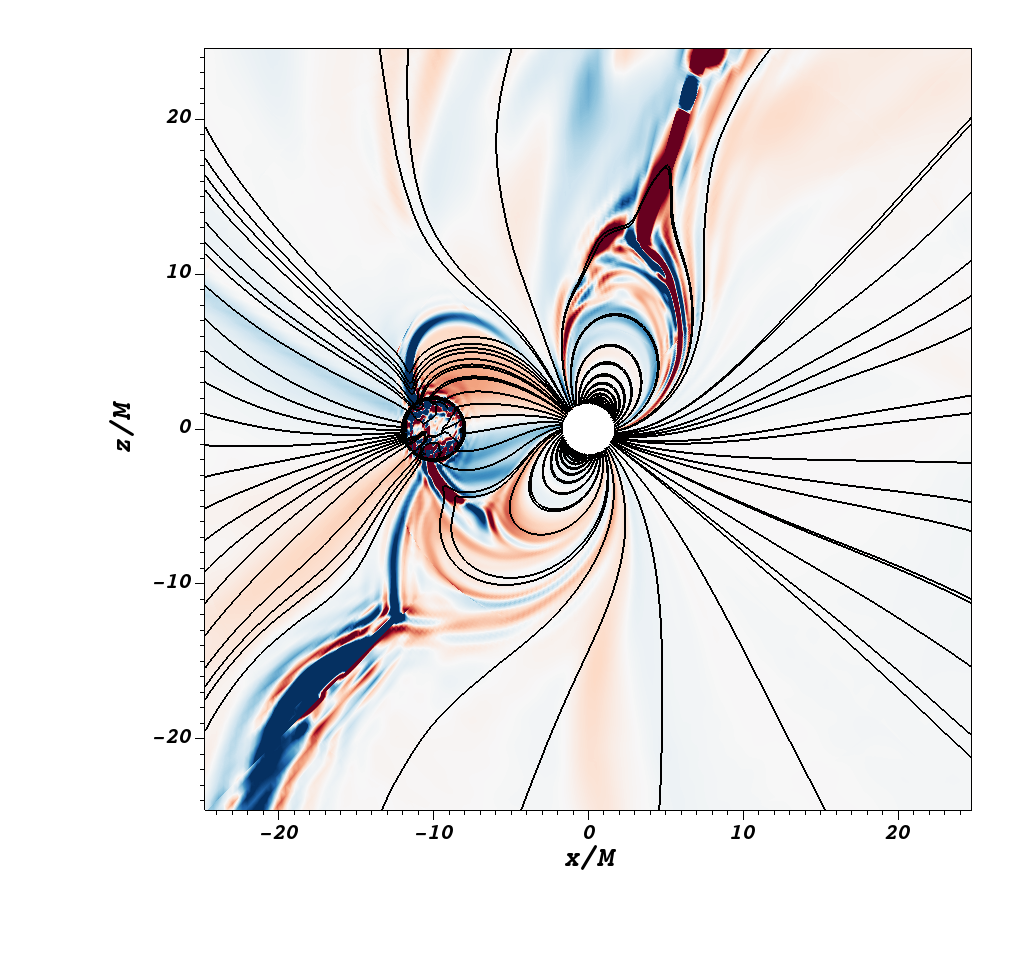}
\includegraphics[scale=0.163]{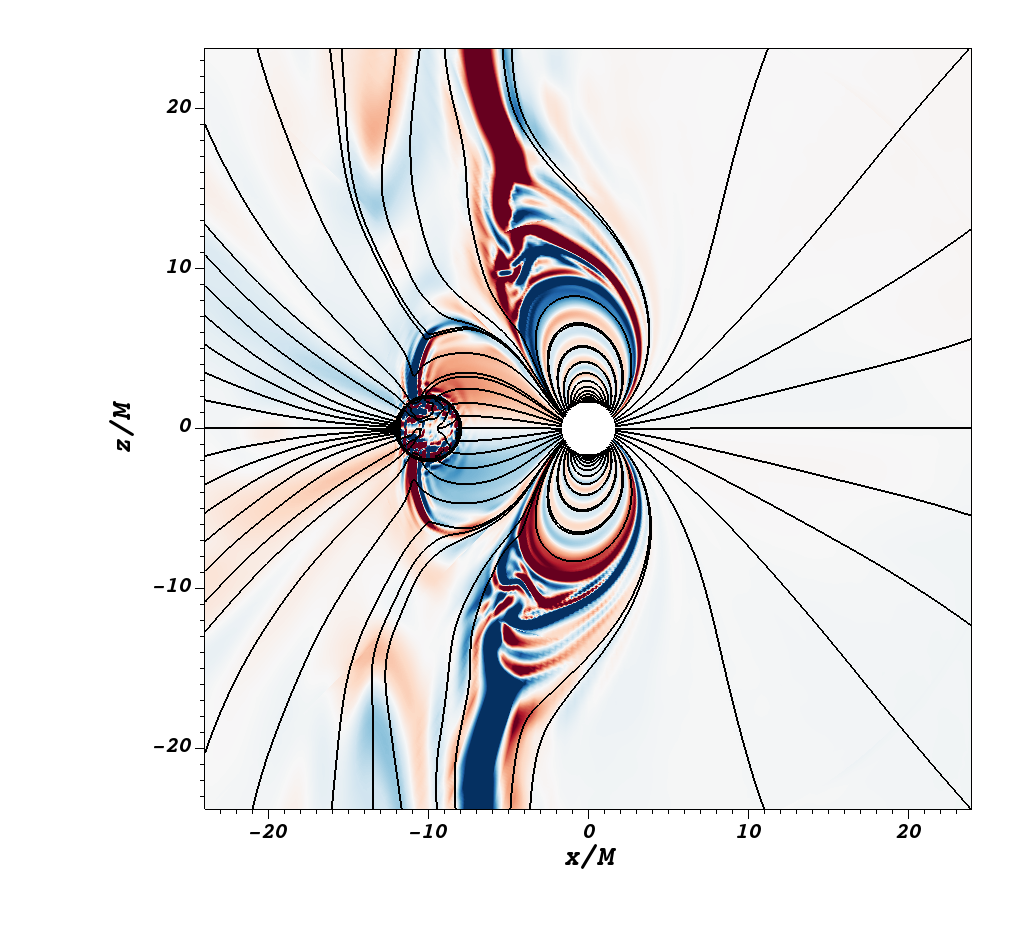}
    \caption{Electric currents for misaligned BHNS binaries in a circular orbit at an orbital separation $r_o = 10 M_{\rm BH}$. 
    Parallel electric currents normalized by $\Omega_o B / 2\pi$ are shown in color scales, along with magnetic field lines projected onto the co-plane.
    The circle centered at the origin depicts the NS, and the thick black circle (located at $x <0$) shows the BH horizon.
    Top panel: snapshots of a $\chi = 15\degree$ misaligned magnetosphere at $2.5$, $2.75$ and $3$ orbits. (An animated version of the sequence can be found \href{https://drive.google.com/drive/folders/1CDfiMedswq6ArcUC-NMt4EB0W6DGsmbq?usp=sharing}{here}).
    Bottom panel: (from left to right) magnetospheres with misalignment $\chi = 30\degree$, $\chi = 60\degree$ and $\chi = 90\degree$, after $3$ orbital periods.
	}
	\label{fig:misaligned_co-plane}
\end{figure*}

The phase dependency is illustrated by the sequence of snapshots presented in the top panel of Fig.~\ref{fig:misaligned_co-plane} for a $\chi = 15\degree$ misaligned magnetosphere, showing parallel currents and projected magnetic field lines on the co-plane. 
Analogous plots (bottom panel) are also presented for the solutions at $3$ orbits for other inclination angles of $\chi = \{ 30\degree, 60\degree, 90\degree \}$. 
Although these solutions are intrinsically three-dimensional and very complex in structure, the co-plane view of Fig.~\ref{fig:misaligned_co-plane} allows us to appreciate their most relevant aspects: (i) there are strong CSs wriggling about the dipole's equators, similar to those seen in \cite{carrasco2020} (cf. their figure 11) and also reminiscent to those of misaligned pulsars; (ii)  magnetic reconnections take place at these CSs; (iii) there are very strong electric currents threading the BH horizon.

\begin{figure}%[!ht]
\centering{
\includegraphics[scale=0.22]{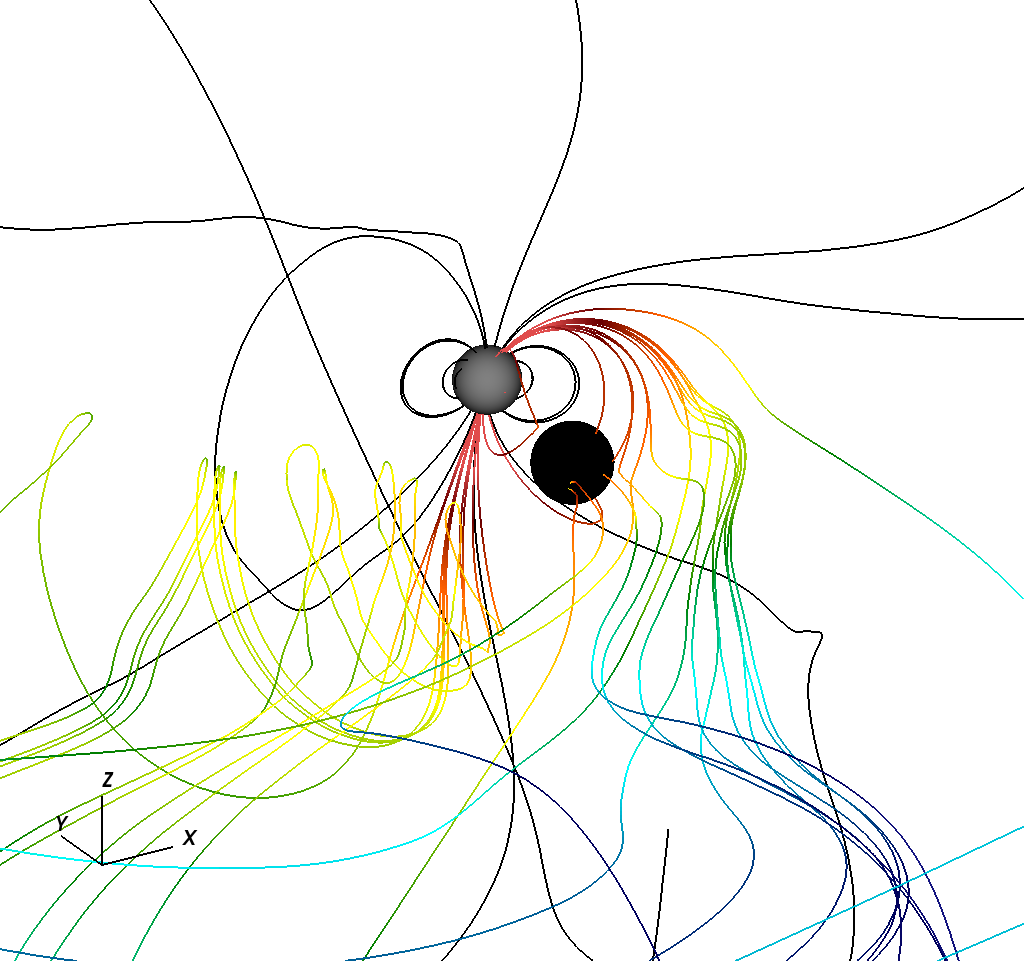}
\caption{Misaligned magnetosphere for a BHNS binary in a circular orbit at an orbital separation $r_o = 10 M_{\rm BH}$ and inclination $\chi = 15\degree$.  Representative magnetic field lines of the solution at $2.75$ orbits are shown, with colors indicating their strength (in logarithmic scale), together with some other reference lines in solid black. The black and grey spheres denote the BH horizon and NS surface, respectively. 
} 
 \label{fig:B1-inclined}}
\end{figure}

The lack of symmetry across the orbital plane makes the previous X-point reconnections less likely to occur and, instead, a larger fraction of the Poynting flux is carried by large amplitude Alfv{\'e}n waves propagating along magnetic field lines that are anchored to the NS. There are, thus, fewer and smaller plasmoids being produced. Figure \ref{fig:B1-inclined} illustrates such disturbances propagating on a $\chi=15\degree$ misaligned magnetosphere at a particular instant along the binary trajectory\footnote{An animation showing the dynamics can be found at this \href{https://drive.google.com/drive/folders/1CDfiMedswq6ArcUC-NMt4EB0W6DGsmbq?usp=sharing}{link}.}. As it can be seen, they resemble the poloidal discontinuities discussed in Sec.~\ref{sec:aligned}, which are produced in the vicinity of the BH due to a combination of its intense gravitational field and the orbital motion. 

This mechanism would suggest a potentially different EM signal as compared to the aligned settings. For instance, analogous large amplitude Alfv{\'e}n wave-packages were linked to fast radio burst originating in magnetars, modeled as coherent radio emission that arises from the decay of such disturbances as they move outwards \cite{kumar2020}. 
%Their shape depend, as mentioned above, on the specifics of the magnetic field lines threading the BH at the particular phase of the binary orbit\footnote{An animation showing the dynamics can be found at this \href{https://drive.google.com/drive/folders/1CDfiMedswq6ArcUC-NMt4EB0W6DGsmbq?usp=sharing}{link}.}. 

%Figure \ref{fig:B1-inclined} illustrates the magnetosphere of a $\chi=15\degree$ misalignment for two instances along the binary trajectory, after $2.75$ and $3$ orbits. In the first one (left panel), the BH is located perpendicular to the misalignment plane of the dipole ($x-z$); whereas in the second plot (right panel), the BH lays at this particular plane. These plots show representative magnetic field lines for each configuration, focusing on the bending effect induced by the strong BH curvature.

%%%%%%%%%%%%%%%%%%%%%%%%%%%%%%%%%%%%%%%%%%%%%%
%%%%%%%%%%%%%%%%%%%%%%%%%%%%%%%%%%%%%%%%%%%%%%

%%%%%%%%%%%%%%%%%%%%%%%%%%%%%%%%%%%%%%%%%%%%%%
\begin{figure}%[!ht]
\centering{
\includegraphics[scale=0.33]{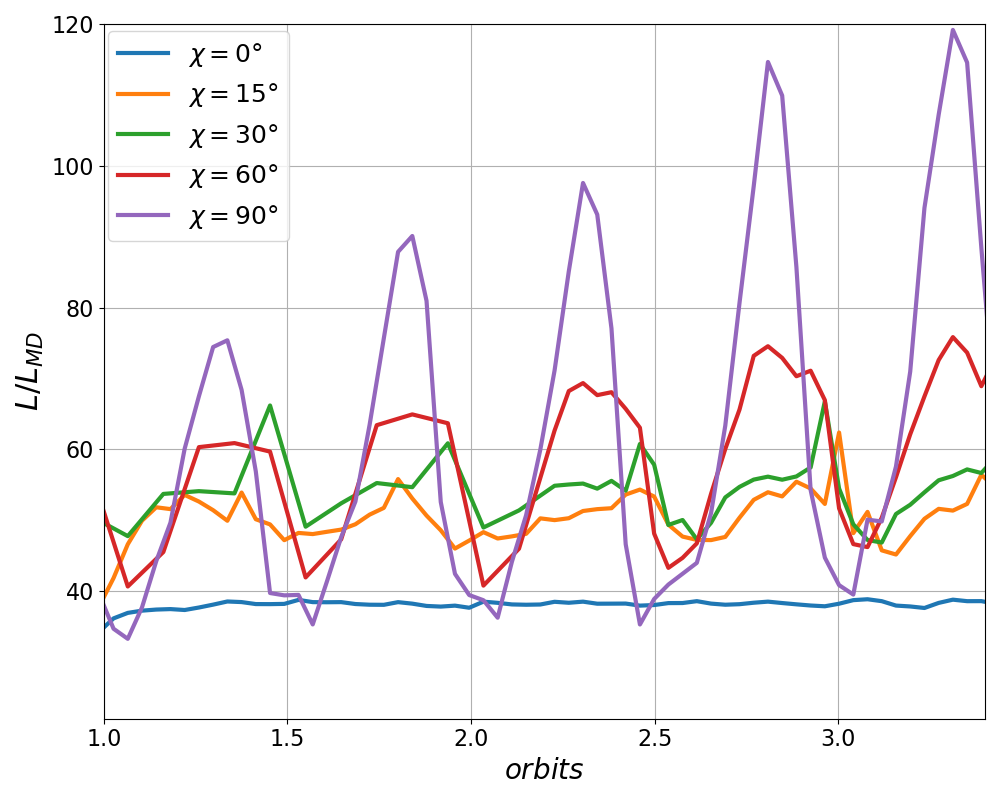}
\caption{EM luminosity of misaligned BHNS binaries in a circular orbit at an orbital separation $r_o = 10 M_{\rm BH}$. Poynting fluxes, integrated at $r=40M_{\rm BH}$, 
shown as a function of time.
}
 \label{fig:Lchi}}
\end{figure}

%%%%%%%%%%%%%%%%%%%%%%%%%%%%%%%%%%%%%%%%%%%%%%
The EM luminosity, measured on a fixed distant surface from the BH, is displayed in Fig.~\ref{fig:Lchi} as a function of time for several misalignment angles $\chi$.
It shows again the phase-dependent character of the misaligned solutions, here regarding the extraction of orbital energy by means of the surrounding plasma.
It is likely that the enhancement of the luminosity (at specific moments along the orbit) relative to the aligned case arises due to higher strength of the magnetic field lines threading the horizon when the inclination of the dipole points towards the BH. 
At this orbital separation, we find an increase (at least\footnote{We note that there is a gradual increase on the peak values with time, especially for larger misalignment angles. This could be a spurious numerical effect (associated with, e.g., lack of resolution) or it could be indicating a longer relaxation timescale for these configurations.}) by a factor of $\sim2$ for the orthogonal case (i.e., $\chi=90\degree$).

%%%%%%%%%%%%%%%%%%%%%%%%%%%%%%%%%%%%%%%%%%%%%%
%%%%%%%%%%%%%%%%%%%%%%%%%%%%%%%%%%%%%%%%%%%%%%

\subsection{Implication to observations}\label{sec3-5}

The magnetosphere generated around the BH and NS can be a source for the EM emission in a wide variety of wavelengths (e.g.,~\cite{wada2020}). We here consider possible EM emissions assuming that the magnetic field strength at the NS pole is $B_p=10^{12}$\,G, paying primary attention to radio waves. 

For radio waves to be emitted from the source, the angular frequency has to be 
higher than the plasma cut-off frequency~\cite{Lightman}
\begin{eqnarray}
 \omega_p=\sqrt{{4\pi \rho_e e \over m_e}}.
\end{eqnarray}
Our results show that the charge 
number density is
\begin{eqnarray}
 n_e&=&f_e {B \Omega_o \over 2\pi c e} \nonumber \\
 &\approx &1.1 \times 10^{10} f_e
 \left({B \over 10^9\,{\rm G}}\right)
 \left({\Omega_o \over 10^3\,{\rm rad/s}}\right)\,{\rm cm^{-3}},
\end{eqnarray}
where $B$ is the local magnetic-field strength and $f_e$ is a numerical 
factor of $\sim 1$--$20$: cf.~Fig.~\ref{fig:co-plane}. 
Thus the value for the cut-off frequency, $\nu_p=\omega_p/(2\pi)$, is
\begin{eqnarray}
 \nu_p \approx 940  f_e^{1/2}
 \left({B \over 10^9\,{\rm G}}\right)^{1/2} 
 \left({\Omega_o \over 10^3\,{\rm rad/s}}\right)^{1/2}\,{\rm MHz}.
\end{eqnarray}
Figure~\ref{fig:co-plane} shows that $f_e$ is larger than unity for the regions 
of dense plasma, and hence, radio waves with the frequency $\nu \alt 1$\,GHz 
are not likely to be efficiently emitted from a region of strong magnetic fields with $B\agt 10^{-3}B_p(=10^9\,{\rm G}$ for $B_p=10^{12}$\,G), which is found in the light cylinder of $r \alt c\Omega_o^{-1}$. 

By contrast, outside the light cylinder, the magnetic-field strength becomes weaker than $10^{-3}B_p$. Even in such region, a CS of the spiral pattern illustrated in Fig.~\ref{fig:B1-CS} is generated.
Such region can be a source for the strong radio emission with the frequency 
less than 1\,GHz. Assuming that a fraction $\epsilon_{\rm eff}$ of the total 
luminosity goes into the radio emission, the spectral flux density at $\nu=1$\,GHz is written as
\begin{eqnarray}
 F_\nu &\approx& 2.1\,{\rm mJy}
 \left({L \over 10^{42}\,{\rm erg/s}}\right)
 \left({\epsilon_{\rm eff} \over 10^{-4}}\right)
 \left({D \over 200\,{\rm Mpc}}\right)^{-2} \nonumber \\
&& \times \left({\nu \over 1\,{\rm GHz}}\right)^{-1},
 \end{eqnarray}
where we supposed the case in which the BH has a small spin and the 
binary is in a close orbit, that can result in the luminosity of $10^{42}\,{\rm erg/s}$. 
However, as already mentioned, the luminosity can be enhanced by two orders of magnitude if the BH spin is high $\agt 0.9$, and hence, a close circular orbit is allowed. Here, we note that the typical radio emission efficiency of pulsars is $10^{-5}$--$10^{-4}$ (see, e.g., Fig.~10 of~\cite{wada2020}). 
With the Square Kilometer Array, whose sensitivity is $\sim 1$\,mJy 
for $0.35$--$1.05$\,GHz (SKA1-MID band1) and 0.95--1.76\,GHz (SKA1-MID band2)~\cite{SKA1,SKA2}, radio waves from BHNSs in close orbits with a 
distance of $\alt 200$\,Mpc may be detected in the future, in particular for 
the binaries of rapidly spinning BHs. 

As indicated by the recent observational results~\cite{Abdo2013,Caraveo2014}, 
strong hard X-rays and gamma-rays are likely to be emitted 
from the magnetosphere of BHNSs in close orbits with a high efficiency. 
However, for the extra-galactic sources with the distance of $D\agt 100$\,Mpc, 
the detection for them is not very optimistic even if the EM luminosity is 
as high as $10^{44}$\,erg/s (e.g., see the estimation in~\cite{wada2020}). 
Even if the BH spin is moderately high, e.g., $\sim 0.9$, for which the 
total maximum EM luminosity can be $\sim 10^{44}(B_p/10^{12}\,{\rm G})^2\,{\rm erg/s}$, the detection by the telescopes currently working such as Swift Alert Telescope and {\em Fermi} Gamma-ray burst Monitor is possible only for the case that the magnetic field strength at the NS pole is very high $\agt 10^{13}$\,G and the emission efficiency in X-rays and gamma-rays is close to unity.

%%%%%%%%%%%%%%%%%%%%%%%%%%%%%%%%%%%%%%%%%%%%%%%%%%%%%%%%%%%%%%%%%%%%%%%%%%%%%%%%%%%%%%%%%%%%%%%%%%%%%% CONCLUSIONS

\section{Conclusions}

In this paper, we have numerically studied the common magnetosphere of an NS in circular orbits around a BH. We constructed the magnetosphere assuming the NS to be well approximated by a perfectly conducting surface endowed with a dipole magnetic field, moving through an FF plasma on a fixed Kerr background.  Although the spacetime geometry is only partially described this way, we expect it to be a good approximation that captures the main phenomena linked to the curvature near the BH.
Such simplified approach has allowed us to analyze the magnetospheric response of these binary systems in great detail and to explore several relevant configurations and parameters.

We found a few generic features of these magnetospheres, present in all of the configurations explored in this work. The most noticeable one is the development of strong CSs that initiate near the BH horizon and extend to large distances, typically following an spiral arm pattern. 
There is also an electric circuit forming along the flux-tube that connects both compact objects, which can be recognized as an approximate realization of the UI effect. 
Additionally, we discovered that the relative motion of the BH respect to the surrounding magnetized plasma produces large twists for a set of magnetic field lines threading the horizon, leading to magnetic reconnections in its vicinity. 
The details of all these features, of course, depend on the specific setting and, especially, on the orientation of the magnetic and orbital axes.
For instance, in the aligned irrotational BHNS binaries studied in Sec.~III-A, X-point reconnections take place immediately to one side of the BH horizon, releasing large-scale plasmoids that carry away most of the EM energy by the Poynting fluxes. By contrast, the lack of symmetry on misaligned configurations make reconnections less likely and a larger fraction of the EM energy is instead transported outwards by large-amplitude Alfv{\'e}n disturbances. 
These two mechanisms, not accounted within the UI model, are responsible for allowing the EM energy to escape from the near-region and be transported to longer distances (where the luminosity is being computed). 

All the above properties can be linked to various EM emissions, as it is the case in the context of isolated pulsar magnetospheres. The electric currents flowing in the flux-tube has been associated with high-energy emissions, either thermal X-rays from Joule heating (e.g., \cite{palenzuela2013linking}) or nonthermal gamma-rays from synchrotron/curvature radiation \cite{mcwilliams2011}; also, this region can generate thermal emissions from relativistic particles bombarding and heating of the NS surface (e.g., \cite{mcwilliams2011, lockhart2019x}); even fast radio bursts from coherent curvature radiation were proposed to originate on this region in the late inspiral phase of BNS systems \cite{wang2016}.
On the other hand, both large-scale plasmoids of aligned settings and the sharp Alfv{\'e}n disturbances found in misaligned cases, which are associated with the binary motion, can be potential sources of radio emissions (e.g.,\cite{kumar2020, most2020, east2021}). 
And finally, the spiral CSs are arguably the most promising source of detectable precursor EM signals for BHNS binaries in close orbits. 

As estimated in Sec.~\ref{sec3-5}, for the most favorable scenarios with high BH spins $\gtrsim 0.9$, with which the EM luminosity can reach $L\sim 10^{44-46} \, [B_p/ 10^{12}{\rm G}]^2  \, {\rm erg/s}$, radio waves could be detected by forthcoming facilities like the Square Kilometer Array at distances of $\lesssim 200$\,Mpc, assuming typical radio emission efficiencies of pulsars. On the other hand, in the high-energy band, albeit strong, the signals are less likely to be detectable by the current instruments at distances $\gtrsim 100$\,Mpc, unless the magnetic field strength is as high as $B_p=10^{13}$\,G.

%%%%%%%%%%%%%%%%%%%%%%%%%%%%%%%%%%%%%%%%%%%%%%%%%%%%%%%%%%%%%%%%%%%%%%%%%%%%%%%%%%%%%%%%%%%%%%%%%%%%%%%%%%%%%%%%%%%%% AGRADECIMIENTOS

\section{Acknowledgments}

We would like to thank Luis Lehner for helpful discussions during the realization of this work. Numerical computations were performed on a Yamazaki cluster at Max Planck Institute for Gravitational Physics at Potsdam and a Sakura cluster at Max-Planck Computing and Data Facility.
This work was in part supported by Grants-in-Aid for Scientific Research (Nos. 16H02183 and 20H00158) of Japanese MEXT/JSPS. 

%%%%%%%%%%%%%%%%%%%%%%%%%%%%%%%%%%%%%%%%%%%%%%%%%%%%%%%%%%%%%%%%%%%%%%%%%%%%%%%%%%%%%%%%%%%%%%%%%%%%%%%%%%%%%%%%%%%%% APENDICES

%%%%%%%%%%%%%%%%%%%%%%%%%%%%%%%%%%%%%%%%%%%%%%%%%%%%%%%%%%%%%%%%%%%%%%%%%%%%%%%%%%%%%%%%%%%%%%%%%%%%%%%%%%%%%%%%%%%%%%%%%%%%%%%%%%%%%%%%%%%%%%%%%%%%%%%%%%%%%%%%%%%
%%%%%%%%%%%%%%%%%%%%%%%%%%%%%%%%%%%%%%%%%%%%%%%%%%%%%%%%%%%%%%%%%%%%%%%%%%%%%%%%%%%%%%%%%%%%%%%%%%%%%%%%%%%%%%%%%%%%%%%%%%%%%%%%%%%%%%%%%%%%%%%%%%%%%%%%%%%%%%%%%%%
%%%%%%%%%%%%%%%%%%%%%%%%%%%%%%%%%%%%%%%%%%%%%%%%%%%%%%%%%%%%%%%%%%%%%%%%%%%%%%%%%%%%%%%%%%%%%%%%%%%%%%%%%%%%%%%%%%%%%%%%%%%%%%%%%%%%%%%%%%%%%%%%%%%%%%%%%%%%%%%%%%%

\bibliographystyle{unsrt} %plain 
\bibliography{FFE}

%%%%%%%%%%%%%%%%%%%%%%%%%%%%%%%%%%%%%%%%%%%%%%%%%%%%%%%%%%%%%%%%%%%%%%%%%%%%%%%%%%%%%%%%%%%%%%%%%%%%%%%%%%%%%%%%%%%%%%%%%%%%%%%%%%%%%%%%%%%%%%%%%%%%%%%%%%%%%%%%%%

\end{document}